\documentclass[10pt,letterpaper,twocolumn,english,journal]{IEEEtran}
\usepackage[T1]{fontenc}
\usepackage{color}
\usepackage{array}
\usepackage{float}
\usepackage{multirow}
\usepackage{amsmath}
\usepackage{graphicx}

\makeatletter

\providecommand{\tabularnewline}{\\}
\floatstyle{ruled}
\newfloat{algorithm}{tbp}{loa}
\providecommand{\algorithmname}{Algorithm}
\floatname{algorithm}{\protect\algorithmname}

\ifCLASSOPTIONcompsoc
\usepackage[caption=false,font=normalsize,labelfont=sf,textfont=sf]{subfig}
\else
\usepackage[caption=false,font=footnotesize]{subfig}
\fi

\usepackage{cite}
\usepackage{url}
\usepackage{algorithmic}
\usepackage{algorithm}
\usepackage{caption}
\usepackage{diagbox}

\makeatother

\usepackage{babel}
\begin{document}
\title{Fair Computation Offloading for RSMA-Assisted Mobile Edge Computing
Networks}
\author{Ding Xu,~\IEEEmembership{Senior Member,~IEEE}, Lingjie Duan,~\IEEEmembership{Senior Member,~IEEE},
Haitao Zhao,~\IEEEmembership{Senior Member,~IEEE}, and Hongbo Zhu,~\IEEEmembership{Member,~IEEE}\thanks{Ding Xu is with the Jiangsu Key Laboratory of Wireless Communications,
Nanjing University of Posts and Telecommunications, Nanjing 210003,
China (E-mail: xuding@ieee.org). He is also with the Pillar of Engineering
Systems and Design, Singapore University of Technology and Design,
Singapore 487372, Singapore.}\thanks{Lingjie Duan is with the Pillar of Engineering Systems and Design,
Singapore University of Technology and Design, Singapore 487372, Singapore
(E-mail: lingjie\_duan@sutd.edu.sg).}\thanks{Haitao Zhao and Hongbo Zhu are with the Jiangsu Key Laboratory of
Wireless Communications, Nanjing University of Posts and Telecommunications,
Nanjing 210003, China (E-mail: zhaoht@njupt.edu.cn; zhuhb@njupt.edu.cn).}}
\maketitle
\begin{abstract}
Rate splitting multiple access (RSMA) provides a flexible transmission
framework that can be applied in mobile edge computing (MEC) systems.
However, the research work on RSMA-assisted MEC systems is still at
the infancy and many design issues remain unsolved, such as the MEC
server and channel allocation problem in general multi-server and
multi-channel scenarios as well as the user fairness issues. In this
regard, we study an RSMA-assisted MEC system with multiple MEC servers,
channels and devices, and consider the fairness among devices. A max-min
fairness computation offloading problem to maximize the minimum computation
offloading rate is investigated. Since the problem is difficult to
solve optimally, we develop an efficient algorithm to obtain a suboptimal
solution. Particularly, the time allocation and the computing frequency
allocation are derived as closed-form functions of the transmit power
allocation and the successive interference cancellation (SIC) decoding
order, while the transmit power allocation and the SIC decoding order
are jointly optimized via the alternating optimization method, the
bisection search method and the successive convex approximation method.
For the channel and MEC server allocation problem, we transform it
into a hypergraph matching problem and solve it by matching theory.
Simulation results demonstrate that the proposed RSMA-assisted MEC
system outperforms current MEC systems under various system setups.
\end{abstract}


\section{Introduction}

\bstctlcite{IEEEexample:BSTcontrol}

In recent years, the rapid development of mobile communications has
brought various mobile applications, such as augmented reality, virtual
reality, and Internet of Things (IoT) \cite{javed2018internet}. However,
it is quite difficult for devices with limited computation capacity
to execute these applications locally. To tackle the computation bottleneck
of devices, mobile edge computing (MEC) has been recognized as a promising
solution \cite{abbas2018mobile}. Specifically, by deploying MEC servers
at the edge of mobile networks, devices can offload their computation
intensive tasks to MEC servers for processing and computing. Since
usually MEC servers can be much more powerful than devices, the task
execution latency can be largely reduced and the computation offloading
rate can be significantly increased.

To fully utilize the computation power of MEC servers, the high efficiency
of computation offloading is desired. Thus, it is important to design
high-efficiency uplink transmission schemes. Due to the shortage of
spectrum resources, it is difficult for conventional orthogonal multiple
access (OMA) transmission schemes such as frequency division multiple
access (FDMA) and time division multiple access (TDMA) to handle the
offloading of computation intensive tasks \cite{xu2021sum}. In this
regard, non-orthogonal multiple access (NOMA) \cite{zheng2020achievable},
which is a promising technique to support multiple transmissions on
the same spectrum band, can be adopted as an efficient uplink transmission
scheme in MEC systems. Specifically, different devices offload their
task data on the same channel using the superposition coding technique,
and the MEC server can partially remove interference among devices
by using the successive interference cancellation (SIC) technique.

In addition to NOMA, rate splitting multiple access (RSMA) is another
novel transmission scheme that uses the rate-splitting technique \cite{mao2022rate}.
RSMA can be adopted both in the downlink and uplink transmissions,
and is slightly different in these two transmissions. In the downlink,
each device message is split into a common part and a private part,
and the common parts of all devices are combined into a single common
message, while each receiver first decodes the common message and
then adopts SIC to decode its private message. In the uplink RSMA,
each device message is split into two sub-messages, and the sub-messages
of all devices are decoded using SIC at the receiver. In contrast
to NOMA where only a single message of each device can be optimized,
both two sub-messages of each device such as their transmit powers
and SIC decoding order can be optimized in RSMA. Thus, RSMA generalizes
NOMA in the uplink transmission, has higher design flexibility, and
can achieve higher system performance. Motivated by this, RSMA can
be applied in MEC systems to further improve the computation offloading
efficiency by flexibly controlling the interference among devices
through message splitting.

\subsection{Related Works}

As an efficient transmission scheme, NOMA has been used to improve
various performance metrics of MEC systems, such as execution latency
\cite{gan2023delay}, energy consumption \cite{wang2023joint}, computation
offloading rate \cite{liu2020resource,xu2024device}, system success
probability \cite{chen2024performance}, and energy efficiency \cite{yu2023computation}.
Since multiple devices compete to share the same communication and
computation resources to offload their computation tasks in NOMA-assisted
MEC systems, fairness among devices is also a significant research
topic that has drawn much attention. The max-min fairness is a well-known
qualitative fairness measure among many fairness measures. Specifically,
max-min fairness is achieved when any device's performance cannot
be increased without sacrificing another device's performance which
is already worse than the previous one. Thus, the max-min fairness
has been applied in NOMA-assisted MEC systems widely \cite{fang2020optimal,kumar2023max}.

Meanwhile, RSMA has been researched in both the downlink \cite{yang2021optimization,zhou2022rate,park2023rate,van2023evolutionary}
and the uplink transmission scenarios \cite{tegos2022performance,yang2022sum,abbasi2023transmission}
in wireless networks. Particularly, as were shown in \cite{tegos2022performance,yang2022sum,abbasi2023transmission},
RSMA provides more design flexibility via message splitting and thus
can achieve higher uplink transmission performance than uplink NOMA.
Due to the advantages of uplink RSMA compared to uplink NOMA as shown
in \cite{tegos2022performance,yang2022sum,abbasi2023transmission},
RSMA can also be adopted in MEC systems to further improve the computation
offloading performance.

To our best knowledge, the work in \cite{han2020rate} was the first
to adopt RSMA in MEC systems, and designed an unmanned aerial vehicle
(UAV) aided MEC system with a UAV and two access points (APs), where
the UAV was assumed to use RSMA for transmitting the collected data
to the two APs for computing. A joint rate splitting scheme to minimize
the UAV energy consumption was proposed, and the UAV position was
designed in \cite{han2020rate}. The work in \cite{reifert2023rate}
extended the work in \cite{han2020rate} to support multiple UAVs.
Specifically, the work in \cite{reifert2023rate} considered a downlink
MEC system with multiple UAVs acting as MEC servers and multiple base
stations (BSs) connecting to a central cloud, and adopted RSMA for
transmitting messages from the UAVs and the BSs to users. The weighted
sum-rate of the system was maximized based on the techniques of discrete
relaxation and fractional programming. In \cite{diamanti2024delay},
concurrent offloading was supported by RSMA in a multi-server MEC
system, and the sum delay was minimized by optimizing the task assignment
ratios, rate and transmit power allocation, and computing resources.

Note that in \cite{han2020rate,reifert2023rate,diamanti2024delay},
the task was assumed to be divisible flexibly to multiple parts and
offloaded to multiple MEC servers adopting the downlink RSMA technique.
However, when there is strong dependence in the task data, such as
the data for face recognition, the task will be indivisible. In addition,
due to the scarce spectrum resource, it is hard to allocate unique
spectrum resource to each task. Thus, uplink RSMA, which can allow
multiple tasks to share the same spectrum resource for offloading,
is more promising than downlink RSMA when the spectrum resource is
short of supply \cite{mao2022rate}. In this context, the works in
\cite{liu2022rate,chen2023rate} applied uplink RSMA in MEC systems.
Specifically, the work in \cite{liu2022rate} considered to use RSMA
to not only protect the task offloading performance of a primary user
(PU) but also maximize the successful computation probability of a
secondary user (SU) in an MEC system. The work in \cite{chen2023rate}
extended the work in \cite{liu2022rate} to consider a group of center
users and edge users and propose to pair one center user and one edge
user to form a pair of PU and SU. The successful computation probability
performance of various user pairing schemes was characterized and
derived in closed form.

\subsection{Motivation and Contributions}

Because uplink RSMA has higher design flexibility than uplink NOMA,
the computation offloading performance of RSMA-assisted MEC systems
can be significantly improved as compared to NOMA-assisted MEC systems.
However, researches on RSMA-assisted MEC systems are still at the
infancy, and many design issues have not been addressed yet.
\begin{itemize}
\item First, current works on uplink RSMA-assisted MEC systems such as \cite{liu2022rate,chen2023rate}
only considered two users, and how to efficiently design offloading
policies with multiple users in an RSMA group remains unknown. The
challenge of this issue is how to coordinate different users for efficient
computation offloading, including the computing frequency allocation,
the transmit power allocation and the SIC decoding order optimization.
\item Second, if multiple MEC servers are available, we can assign a proper
MEC server to each RSMA group to enhance the computation offloading
performance, whereas the policy of MEC server allocation is unexplored
in existing works on uplink RSMA-assisted MEC systems.
\item Third, if more than one channel is available, multiple RSMA groups
can be formed to perform offloading, and the device grouping problem
also known as the channel allocation problem needs to be solved for
better computation offloading performance, which is ignored in existing
studies. Note that the MEC server allocation and the channel allocation
are usually coupled, i.e., the devices allocated with the same channel
shall be also allocated with the same MEC server for interference
mitigation. The challenge of the second and the third issues is how
to optimize the coupled and discrete MEC server allocation and channel
allocation variables.
\item Fourth, user fairness, which is an important design objective, is
not of concern in the existing works on RSMA-assisted MEC systems.
The challenge of this issue is how to strike a balance among the performance
of different devices.
\end{itemize}

The above remaining issues motivate this work. This paper considers
an RSMA-assisted MEC system with multiple MEC servers, multiple channels
and multiple IoT devices, and investigates the joint communication
and computation resource allocation (including the MEC server allocation
and the channel allocation) problem, targeting achieving a max-min
fairness computation offloading, i.e., the minimum computation offloading
rate (MCOR) among all devices is maximized. 

It is worth noting that the scenario of multiple MEC servers, multiple
channels and multiple IoT devices with the assistance of RSMA has
not been researched yet. Particularly, for the first design issue,
we first resort to the analytical method to simplify the design, and
then heuristically use the alternating optimization method, the bisection
search method and the SCA method to jointly optimize the transmit
power allocation and the SIC decoding order. For the second and third
design issues, we first transform the joint MEC server allocation
and channel allocation problem into a hypergraph matching problem,
and then rely on the matching theory to derive an efficient algorithm
to solve the problem. For the fourth design issue, we adopt the max-min
fairness in the design, and aim to maximize the MCOR to make a tradeoff
among the performance of different devices.

The main contributions of this paper are:
\begin{itemize}
\item We investigate an RSMA-assisted MEC system with multiple MEC servers,
channels and IoT devices, and propose to divide multiple devices into
several groups, each of which can be allocated with one channel and
formed into an RSMA group to offload task data to an allocated MEC
server for processing. The problem of optimizing the channel allocation,
the MEC server allocation, the time allocation, the computing frequency
allocation, the transmit power allocation and the SIC decoding order,
aiming at maximizing the MCOR among all devices is investigated.
\item The formulated problem is an intractable non-convex mixed integer
programming problem, whose optimal solution is difficult to find.
Thus, we develop an efficient algorithm to obtain a suboptimal solution.
Specifically, given the channel allocation and the MEC server allocation,
the problem is transformed to parallel subproblems, each for an MEC
server. By exploring the particular problem structure, we first analytically
derive the time allocation and the computing frequency allocation
as closed-form functions of the transmit power allocation and the
SIC decoding order, then jointly obtain the transmit power allocation
and the SIC decoding order heuristically based on the alternating
optimization method, the bisection search method and the SCA method.
For the channel and MEC server allocation, the problem is a nonlinear
integer problem. We first transform it into a hypergraph matching
problem and then propose a low-complexity matching theory based algorithm
to solve it. Particularly, the channel allocation problem is recast
as a one-to-many matching problem with externalities and solved by
swap matching, while the MEC server allocation problem is recast as
a one-to-many matching problem and solved by the Gale-Shapley algorithm.
\item Simulations demonstrate that the proposed RSMA-assisted MEC system
performs better than current MEC systems such as the NOMA and TDMA
assisted MEC systems, as well as the benchmark RSMA-assisted MEC systems.
\end{itemize}

The rest of this paper is organized as follows. Section \ref{sec:System-Model}
introduces the system model and formulates the MCOR maximization problem.
Section \ref{sec:Solution-for-the} proposes an algorithm to optimize
the time allocation, the computing frequency allocation, the transmit
power allocation and the SIC decoding order. Section \ref{sec:MEC-Allocation-and}
proposes a matching theory based algorithm to optimize the channel
allocation and the MEC server allocation. Section \ref{sec:Simulation-Results}
illustrates simulation results to demonstrate the effectiveness of
the proposed algorithm. Finally, the paper is concluded in Section
\ref{sec:Conclusions}.

\section{System Model and Problem Formulation\label{sec:System-Model}}

\begin{figure}[!t]
\centering\includegraphics[width=0.9\columnwidth]{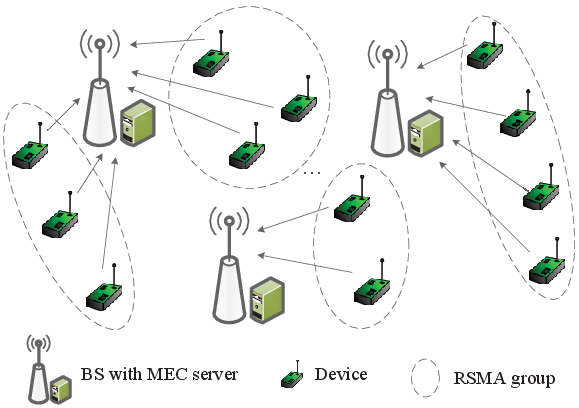}

\caption{System model.\label{fig:System-model}}
\end{figure}

\subsection{System Description\label{subsec:System-Description}}

We consider an MEC system consisting of $K$ IoT devices and $M$
BSs, each of which is equipped with an MEC server, as shown in Fig.
\ref{fig:System-model}. Since each BS is co-located with one MEC
server, we use BS and MEC server interchangeably. We assume that each
device has some task data to be computed, which are indivisible, i.e.,
there is strong dependence over different parts of the task data (such
as face recognition). Due to the limited computation capacity of the
devices such as sensors in IoT, the task data cannot be processed
locally and need to be offloaded to an MEC server for processing and
computing first, and then the MEC server may send the computing results
(such as the face recognition results, or the command and control
data) to the corresponding devices. This scenario is applicable in
practice. For example, devices in the cell edge have multiple BSs
that can be connected to, and each device can choose one BS for task
offloading. In addition, in ultra-dense MEC networks \cite{zhou2024cost},
each device has multiple candidate BSs within the communication range
and thus can choose one BS for task offloading.

The spectrum of interest is divided into $N$ orthogonal channels,
each has a bandwidth $B$. The sets of MEC servers, channels and devices
are denoted by $\mathcal{M},$ $\mathcal{N}$ and $\mathcal{K}$,
respectively. The MEC server $m,$ channel $n,$ and device $k$ are
denoted by $\mathrm{MEC}_{m}$, $\mathrm{CH}_{n}$ and $\mathrm{DVC}_{k}$,
respectively. Delay-sensitive applications are assumed to be carried
by the devices and the task data has to be finished computing within
the time deadline $T$, and each MEC server divides the time into
two phases. In the first phase with time $t_{m}^{\mathrm{o}}$, the
associated devices offload their data to $\mathrm{MEC}_{m}$. Since
the data of each device are indivisible, they can be offloaded to
only one MEC server for computing, i.e. $\sum_{m\in\mathcal{M}}\alpha_{m,k}\leq1,\forall k,$
where $\alpha_{m,k}\in\{0,1\}$ is the MEC server allocation variable.
Specifically, $\alpha_{m,k}=1$ means that $\mathrm{DVC}_{k}$ offloads
its data to $\mathrm{MEC}_{m}$, and $\alpha_{m,k}=0$ indicates otherwise. 

It is also assumed that each device can be allocated with one channel
for data offloading, i.e. $\sum_{n\in\mathcal{N}}\beta_{n,k}\leq1,\forall k,$
where $\beta_{n,k}\in\{0,1\}$ is the channel allocation variable.
Particularly, $\beta_{n,k}=1$ means that $\mathrm{DVC}_{k}$ is allocated
with $\mathrm{CH}_{n}$ for data offloading, and $\beta_{n,k}=0$
indicates otherwise. Besides, in order to eliminate the inter-cell
interference in such a multi-BS communication scenario, one channel
cannot be allocated to the devices associated with different MEC servers,
i.e., $\beta_{n,k}\beta_{n,k^{'}}\alpha_{m,k}\alpha_{m^{'},k^{'}}=0,\forall n,\forall k,\forall k^{'}\neq k,\forall m,\forall m^{'}\neq m.$
Since one channel can be allocated to multiple devices at the same
MEC server, RSMA is proposed to be adopted. Specially, the transmitted
message $s_{k}$ of each $\mathrm{DVC}_{k}$ is split into two sub-messages
$s_{k,1}$ and $s_{k,2}$ , and rate splitting is achieved by allocating
different transmit powers to the two sub-messages, i.e., $p_{k,1}+p_{k,2}\leq P_{k},\forall k,$
where $P_{k}$ is the maximum transmit power of device $k,$ and $p_{k,1}$
and $p_{k,2}$ are the transmit powers of the sub-messages $s_{k,1}$
and $s_{k,2},$ respectively. Thus, we have $s_{k}=\sqrt{p_{k,1}}s_{k,1}+\sqrt{p_{k,2}}s_{k,2},\forall k.$
The total message $y_{m,n}$ received at $\mathrm{MEC}_{m}$ on $\mathrm{CH}_{n}$
can be written as
\begin{align}
y_{m,n} & =\sum_{k\in\mathcal{K}}\alpha_{m,k}\beta_{n,k}\sqrt{h_{m,n,k}}s_{k}+n_{0}\nonumber \\
 & =\sum_{k\in\mathcal{K}}\alpha_{m,k}\beta_{n,k}\sum_{i=1}^{2}\sqrt{h_{m,n,k}p_{k,i}}s_{k,i}+n_{0},
\end{align}
where $h_{m,n,k}$ is the channel gain between $\mathrm{DVC}_{k}$
and $\mathrm{MEC}_{m}$ on $\mathrm{CH}_{n}$, and $n_{0}$ is the
additive white Gaussian noise.

Then, the MEC server adopts SIC to decode the sub-messages of all
associated devices. Let $\pi_{k,i},i\in\{1,2\}$ denote the SIC decoding
order of sub-message $s_{k,i}$ from $\mathrm{DVC}_{k}$. Specifically,
the SIC decoding order of all the devices is denoted by a permutation
$\boldsymbol{\pi}=\{\pi_{k,i},\forall k,\forall i\},$ which belongs
to the set $\Pi$ defined as the set of all possible SIC decoding
orders of all $2K$ sub-messages. Then, the achievable rate of decoding
sub-message $s_{k,i}$ at $\mathrm{MEC}_{m}$ on $\mathrm{CH}_{n}$
is
\begin{align}
 & r_{m,n,k,i}\!=\!B\log_{2}\left(\!\!1\!+\!\frac{h_{m,n,k}p_{k,i}}{\sigma^{2}B\!+\!\sum_{(k^{'},i^{'})\in\mathcal{S}_{m,n,k,i}}h_{m,n,k^{'}}p_{k^{'},i^{'}}}\!\!\right),\label{eq:rate}
\end{align}
where $\sigma^{2}$ is the power spectral density of the noise, and
$\mathcal{S}_{m,n,k,i}=\{(k^{'},i^{'})|\alpha_{m,k^{'}}=\alpha_{m,k},\beta_{n,k^{'}}=\beta_{n,k},\pi_{k^{'},i^{'}}>\pi_{k,i},\forall k^{'},i^{'}\}$
is the set of sub-messages that are decoded after sub-message $s_{k,i}$
at the MEC server $\mathrm{MEC}_{m}$ on the same channel. Since the
message $s_{k}$ of $\mathrm{DVC}_{k}$ consists of sub-messages $s_{k,1}$
and $s_{k,2}$, the computation offloading rate of $\mathrm{DVC}_{k}$
is given by \cite{zheng2020achievable}
\begin{equation}
r_{k}=\frac{t_{m}^{\mathrm{o}}}{T}\sum_{m\in\mathcal{M}}\sum_{n\in\mathcal{N}}\sum_{i=1}^{2}\alpha_{m,k}\beta_{n,k}r_{m,n,k,i}.\label{eq:rate-k}
\end{equation}

In the second phase with time $t_{m}^{\mathrm{c}}$, $\mathrm{MEC}_{m}$
processes and computes the data offloaded from all the associated
devices. Let $f_{k}$ (in offloaded bits per second) denote the computing
frequency allocated for computing the data offloaded by $\mathrm{DVC}_{k}$.
Then, we have $\sum_{k\in\mathcal{K}}\alpha_{m,k}f_{k}\leq F_{m},\forall m,$
where $F_{m}$ is the total computing frequency of $\mathrm{MEC}_{m}$.
Since the offloaded data in the first phase have to be finished computing
at the MEC server at the end of the second time phase, we have $\frac{r_{k}T}{\sum_{m\in\mathcal{M}}\alpha_{m,k}f_{k}}\leq\sum_{m\in\mathcal{M}}\alpha_{m,k}t_{m}^{\mathrm{c}},\forall k.$
Since the computing results of the offloaded data are usually much
smaller than the task data size and the transmit power of the BSs
is much higher than the transmit power of the devices, the time for
sending the results back is neglected.

\subsection{Problem Formulation}

Since all devices compete to share the resources and there exists
interference among devices in the same RSMA group, the interests of
different devices are in conflict. This means that if no measure is
taken, there may exist some devices in inferior channel conditions
whose computation offloading performance will be much worse than other
devices. Such unfair computation offloading performance among devices
in MEC systems is unacceptable and may lead to user complaints. Therefore,
to guarantee the computation offloading performance of each device,
fairness among devices is considered in this work. Currently, there
are two most representative fairness measures, namely max-min fairness
and proportional fairness. Compared to proportional fairness, max-min
fairness can provide fairer resource allocation from the system perspective
\cite{huaizhou2014fairness}. Therefore, max-min fairness is adopted
as the design objective to achieve fairness among devices. Max-min
fairness has been widely used in MEC systems for fair computation
offloading \cite{du2018computation,liu2020max}.

We are interested in achieving max-min fairness for all devices by
maximizing the MCOR among all devices, where the MEC server allocation
$\boldsymbol{\alpha}=\{\alpha_{m,k},\forall m,\forall k\},$ the channel
allocation $\boldsymbol{\beta}=\{\beta_{n,k},\forall n,\forall k\},$
the time allocation $\mathbf{t}^{\mathrm{o}}=\{t_{m}^{\mathrm{o}},\forall m\}$
and $\mathbf{t}^{\mathrm{c}}=\{t_{m}^{\mathrm{c}},\forall m\},$ the
computing frequency allocation $\mathbf{f}=\{f_{k},\forall k\},$
the transmit power allocation $\mathbf{p}=\{p_{k,i},\forall k,\forall i\},$
and the SIC decoding order $\boldsymbol{\pi}$ are jointly optimized.

From the system description in Section \ref{subsec:System-Description},
the objective function defined as the MCOR among all devices is given
by $\min_{k\in\mathcal{K}}r_{k}$, and all the constraints of the
problem have also been presented. Mathematically, the optimization
problem is formulated as
\begin{subequations}
\label{eq:p1}
\begin{alignat}{1}
\max & \:\min_{k\in\mathcal{K}}r_{k}\label{eq:p1-o}\\
\mathrm{s.t.} & \:\sum_{m\in\mathcal{M}}\alpha_{m,k}\leq1,\forall k,\label{eq:p1-c0}\\
 & \:\sum_{n\in\mathcal{N}}\beta_{n,k}\leq1,\forall k,\label{eq:p1-c1}\\
 & \:\beta_{n,k}\beta_{n,k^{'}}\alpha_{m,k}\alpha_{m^{'},k^{'}}=0,\forall n,\forall k,\forall k^{'}\neq k,\nonumber \\
 & \hphantom{\beta_{n,k}\beta_{n,k^{'}}\alpha_{m,k}\alpha_{m^{'},k^{'}}=0,\forall n}\forall m,\forall m^{'}\neq m,\label{eq:p1-c1-1}\\
 & \:p_{k,1}+p_{k,2}\leq P_{k},\forall k,\label{eq:p1-c2}\\
 & \:\sum_{k\in\mathcal{K}}\alpha_{m,k}f_{k}\leq F_{m},\forall m,\label{eq:p1-c4}\\
 & \:\frac{r_{k}T}{\sum_{m\in\mathcal{M}}\alpha_{m,k}f_{k}}\leq\sum_{m\in\mathcal{M}}\alpha_{m,k}t_{m}^{\mathrm{c}},\forall k,\label{eq:p1-c5}\\
 & \:t_{m}^{\mathrm{o}}+t_{m}^{\mathrm{c}}\leq T,\forall m,\label{eq:p1-c6}\\
 & \:\alpha_{m,k}\in\{0,1\},\forall m,\forall k,\label{eq:p1-c7}\\
 & \:\beta_{n,k}\in\{0,1\},\forall n,\forall k,\label{eq:p1-c8}\\
 & \:t_{m}^{\mathrm{o}}\geq0,t_{m}^{\mathrm{c}}\geq0,\forall m,\label{eq:p1-c9}\\
 & \:f_{k}\geq0,\forall k,\label{eq:p1-c10}\\
 & \:p_{k,i}\geq0,\forall k,\forall i,\label{eq:p1-c11}\\
 & \:\boldsymbol{\pi}\in\Pi,\label{eq:p1-c12}\\
\mathrm{o.v.} & \:\boldsymbol{\alpha},\boldsymbol{\beta},\mathbf{t}^{\mathrm{o}},\mathbf{t}^{\mathrm{c}},\mathbf{f},\mathbf{p},\boldsymbol{\pi},
\end{alignat}
\end{subequations}
where \textquoteleft $\mathrm{s.t.}$\textquoteright{} stands for
\textquoteleft subject to\textquoteright{} and \textquoteleft $\mathrm{o.v.}$\textquoteright{}
stands for \textquoteleft optimization variables\textquoteright .
In the following, we give detailed explanations for the constraints
in the problem \eqref{eq:p1}. The constraint \eqref{eq:p1-c0} guarantees
that the data of each device can be offloaded to at most one MEC server
for computing. The constraint \eqref{eq:p1-c1} restricts that each
device can be allocated with at most one channel for data offloading.
The constraint \eqref{eq:p1-c1-1} restricts that one channel can
only be allocated to the devices assigned to the same MEC server.
The constraint \eqref{eq:p1-c2} restricts that the total transmit
power of the two sub-messages of each device cannot exceed the maximum
transmit power. The constraint \eqref{eq:p1-c4} restricts that the
total computing frequency allocated to all the devices cannot exceed
the total computing frequency at the MEC server. The constraint \eqref{eq:p1-c5}
guarantees that the data offloaded to each MEC server have to be finished
computing at the end of the considered time. The constraint \eqref{eq:p1-c6}
guarantees that the time of the two phases cannot exceed the time
deadline. The constraints \eqref{eq:p1-c7} and \eqref{eq:p1-c8}
guarantee that the MEC server allocation and the channel allocation
are either $0$ or $1$, respectively. The constraints \eqref{eq:p1-c9},
\eqref{eq:p1-c10} and \eqref{eq:p1-c11} guarantee non-negative time
allocation, computing frequency and transmit power, respectively.
The constraint \eqref{eq:p1-c12} restricts the SIC decoding order. 

Due to the discrete $\boldsymbol{\alpha}$ and $\boldsymbol{\beta}$,
and the non-convex objective function and the constraint \eqref{eq:p1-c5},
the problem \eqref{eq:p1} is a non-convex mixed integer programming
problem. Thus, it is generally hard to optimally solve the problem
\eqref{eq:p1}. In this regard, we will next develop a four-step algorithm
to obtain a suboptimal solution. Specifically, in the first step,
$\boldsymbol{\beta}$ is initialized by randomly allocating a channel
to each device, and then $\boldsymbol{\alpha}$ is initialized by
randomly allocating an MEC server to each (channel, device) unit.
In the second step, $\mathbf{t}^{\mathrm{o}},$ $\mathbf{t}^{\mathrm{c}},$
$\mathbf{f},$ $\mathbf{p}$ and $\boldsymbol{\pi}$ are optimized
by exploring the particular problem structure and using the alternating
optimization method, the bisection search method and the SCA method.
In the third step, a low-complexity matching theory based algorithm
is proposed to optimize $\boldsymbol{\alpha}$ and $\boldsymbol{\beta}$.
In the final step, the methods used in the second step are adopted
again to optimize $\mathbf{t}^{\mathrm{o}},$ $\mathbf{t}^{\mathrm{c}},$
$\mathbf{f},$ $\mathbf{p}$ and $\boldsymbol{\pi}$. The proposed
algorithms in the second/fourth step and the third step will be discussed
in Section \ref{sec:Solution-for-the} and Section \ref{sec:MEC-Allocation-and},
respectively. The overall proposed algorithm to solve the problem
\eqref{eq:p1} is summarized in Algorithm \ref{alg:0}.

Note that in order to execute the proposed algorithm, a central unit
similar to the one proposed in \cite{ju2022energy,pan2024energy}
that connects with all the BSs through high-speed optical fiber can
be used to coordinate the cooperation among BSs and collect necessary
information for decision making. Specifically, each BS first collects
the channel state information of the devices and sends the information
to the central unit along with other necessary parameters such as
the transmit power budget and the maximum computing frequency. Then,
the central unit executes the proposed algorithm to decide the MEC
server allocation and channel allocation, as well as the time allocation,
computing frequency allocation, power allocation and SIC decoding
order. The decision is then sent from the central unit to all the
BSs for execution.

\begin{algorithm}[!t]
\begin{algorithmic}[1]
\STATE Initialize $\boldsymbol{\beta}$ by randomly allocating a channel to each device, and then initialize $\boldsymbol{\alpha}$ by randomly allocating an MEC server to each (channel, device) unit.
\STATE Obtain $\mathbf{t}_{o},$ $\mathbf{t}_{c},$ $\mathbf{f},$ $\mathbf{p}$ and $\boldsymbol{\pi}$ by the algorithm proposed in Section \ref{sec:Solution-for-the}. 
\STATE Update  $\boldsymbol{\alpha}$ and $\boldsymbol{\beta}$ according to the algorithm proposed in Section \ref{sec:MEC-Allocation-and}.
\STATE Update $\mathbf{t}_{o},$ $\mathbf{t}_{c},$ $\mathbf{f},$ $\mathbf{p}$ and $\boldsymbol{\pi}$ by the algorithm proposed in Section \ref{sec:Solution-for-the}.
\STATE Output: $\boldsymbol{\alpha},$ $\boldsymbol{\beta},$ $\mathbf{t}_{o},$ $\mathbf{t}_{c},$ $\mathbf{f},$ $\mathbf{p}$ and $\boldsymbol{\pi}$.
\end{algorithmic}\caption{Proposed algorithm for solving the problem \eqref{eq:p1}.\label{alg:0}}
\end{algorithm}

\section{Solution To the Problem of Optimizing Time Allocation, Computing
Frequency Allocation, Power Allocation and SIC Decoding Order\label{sec:Solution-for-the}}

In this section, with given MEC server allocation $\boldsymbol{\alpha}$
and channel allocation $\boldsymbol{\beta}$, an efficient algorithm
is designed to solve the problem of optimizing $\mathbf{t}^{\mathrm{o}},$
$\mathbf{t}^{\mathrm{c}},$ $\mathbf{f},$ $\mathbf{p}$ and $\boldsymbol{\pi}$
in the problem \eqref{eq:p1}. By introducing an auxiliary variable
$\eta$, the problem of optimizing $\mathbf{t}^{\mathrm{o}},$ $\mathbf{t}^{\mathrm{c}},$
$\mathbf{f},$ $\mathbf{p}$ and $\boldsymbol{\pi}$ in the problem
\eqref{eq:p1} can be rewritten as 
\begin{subequations}
\label{eq:p1-1}
\begin{alignat}{1}
\max_{\eta,\mathbf{t}^{\mathrm{o}},\mathbf{t}^{\mathrm{c}},\mathbf{f},\mathbf{p},\boldsymbol{\pi}} & \:\eta\\
\mathrm{s.t.} & \:r_{k}\geq\eta,\forall k,\label{eq:p1-1-c1}\\
 & \:\eqref{eq:p1-c2}-\eqref{eq:p1-c6},\eqref{eq:p1-c9}-\eqref{eq:p1-c12}.\nonumber 
\end{alignat}
\end{subequations}
Suppose that $\eta^{*}$ is the optimal $\eta.$ Then, the problem
\eqref{eq:p1-1} with given $\eta$ is feasible if $\eta\leq\eta^{*}$,
and is infeasible if $\eta>\eta^{*}$. Thus, the feasibility of the
problem has two opposite results depending on the value of $\eta.$
In this regard, the bisection search method \cite{convexop2004},
which consists of iteratively bisecting the interval and selecting
the sub-interval that contains the optimal solution, can be conducted
to find the optimal $\eta^{*}$. Specifically, the sub-interval with
larger values of $\eta$ is selected if the problem \eqref{eq:p1-1}
with the bisected $\eta$ is feasible, and the sub-interval with smaller
values of $\eta$ is selected otherwise. In what follows, we focus
on checking the feasibility of the problem \eqref{eq:p1-1} with given
$\eta$, the problem of which is formulated as
\begin{alignat}{1}
\mathrm{Find} & \:\mathbf{t}^{\mathrm{o}},\mathbf{t}^{\mathrm{c}},\mathbf{f},\mathbf{p},\boldsymbol{\pi}\label{eq:p1-2}\\
\mathrm{s.t.} & \:\eqref{eq:p1-c2}-\eqref{eq:p1-c6},\eqref{eq:p1-c9}-\eqref{eq:p1-c12},\eqref{eq:p1-1-c1}.\nonumber 
\end{alignat}
Since MEC allocation and channel allocation have been performed, the
above feasibility-check problem can be decomposed into $M$ subproblems,
each for an MEC server and its associated devices on allocated channels.
Let $\mathcal{K}_{m}$ denote the set of devices associated with $\mathrm{MEC}_{m}$.
Then, the feasibility-check subproblem for $\mathrm{MEC}_{m}$ is
formulated as
\begin{subequations}
\label{eq:p1-3}
\begin{alignat}{1}
\mathrm{Find} & \:\mathbf{t}_{m},\mathbf{f}_{m},\mathbf{p}_{m},\boldsymbol{\pi}_{m}\\
\mathrm{s.t.} & \:p_{k,1}+p_{k,2}\leq P_{k},\forall k\in\mathcal{K}_{m},\label{eq:p1-3-c1}\\
 & \:r_{k}\geq\eta,\forall k\in\mathcal{K}_{m},\\
 & \:\sum_{k\in\mathcal{K}_{m}}f_{k}\leq F_{m},\\
 & \:\frac{r_{k}T}{f_{k}}\leq t_{m}^{\mathrm{c}},\forall k\in\mathcal{K}_{m},\label{eq:p1-3-c4}\\
 & \:t_{m}^{\mathrm{o}}+t_{m}^{\mathrm{c}}\leq T,\label{eq:p1-3-c5}\\
 & \:t_{m}^{\mathrm{o}}\geq0,t_{m}^{\mathrm{c}}\geq0,\\
 & \:f_{k}\geq0,\forall k\in\mathcal{K}_{m},\\
 & \:p_{k,i}\geq0,\forall k\in\mathcal{K}_{m},\forall i,\label{eq:p1-3-c8}\\
 & \:\boldsymbol{\pi}_{m}\in\Pi_{m},\label{eq:p1-3-c9}
\end{alignat}
\end{subequations}
where $\mathbf{t}_{m}=\{t_{m}^{\mathrm{o}},t_{m}^{\mathrm{c}}\},$
$\mathbf{f}_{m}=\{f_{k},\forall k\in\mathcal{K}_{m}\},$ $\mathbf{p}_{m}=\{p_{k,i},\forall k\in\mathcal{K}_{m},\forall i\},$
$\boldsymbol{\pi}_{m}=\{\pi_{k,i},\forall k\in\mathcal{K}_{m},\forall i\}$
is the SIC decoding order of the devices associated with $\mathrm{MEC}_{m}$,
$\Pi_{m}$ is the set of all possible SIC decoding orders of all $2|\mathcal{K}_{m}|$
sub-messages from $|\mathcal{K}_{m}|$ devices, $r_{k}=\frac{t_{m}^{\mathrm{o}}}{T}\sum_{n\in\mathcal{N}}\sum_{i=1}^{2}\beta_{n,k}r_{m,n,k,i},$
and $\mathcal{S}_{m,n,k,i}$ in the expression of $r_{m,n,k,i}$ in
\eqref{eq:rate-k} is rewritten as $\mathcal{S}_{m,n,k,i}=\{(k^{'},i^{'})|\beta_{n,k^{'}}=\beta_{n,k},\pi_{k^{'},i^{'}}>\pi_{k,i},\forall k^{'}\in\mathcal{K}_{m},\forall i^{'}\}$
which represents the set of sub-messages that are decoded after sub-message
$s_{k,i}$ at the MEC server $\mathrm{MEC}_{m}$ on the same channel.
Note that if all the $M$ feasibility-check subproblems are feasible,
then the problem \eqref{eq:p1-1} with given $\eta$ is feasible;
otherwise, the problem \eqref{eq:p1-1} with given $\eta$ is infeasible.

To solve the problem \eqref{eq:p1-3}, the properties of the constraints
with the feasible solution are first presented.

\noindent \textbf{Proposition 1}: The constraint \eqref{eq:p1-3-c4}
can be satisfied with equality by the feasible solution to the problem
\eqref{eq:p1-3}, i.e., $\frac{r_{k}T}{f_{k}}=t_{m}^{\mathrm{c}},\forall k\in\mathcal{K}_{m}.$
\begin{IEEEproof}
If for a $k^{'}\in\mathcal{K}_{m}$, we have$\frac{r_{k^{'}}T}{f_{k^{'}}}<t_{m}^{\mathrm{c}}.$
Then, we can decrease $f_{k^{'}}$ until $f_{k^{'}}=\frac{r_{k^{'}}T}{t_{m}^{\mathrm{c}}},$
such that the constraints of the problem \eqref{eq:p1-3} are still
satisfied.
\end{IEEEproof}
Proposition 1 means that the computing frequency of the MEC server
allocated to each device shall be just enough to compute the offloaded
data.

\noindent \textbf{Proposition 2}: The constraint \eqref{eq:p1-3-c5}
can be satisfied with equality by the feasible solution to the problem
\eqref{eq:p1-3}, i.e., $t_{m}^{\mathrm{o}}+t_{m}^{\mathrm{c}}=T.$
\begin{IEEEproof}
If for a $m,$ we have $t_{m}^{\mathrm{o}}+t_{m}^{\mathrm{c}}<T.$
Then, $t_{m}^{\mathrm{c}}$ can be increased until $t_{m}^{\mathrm{o}}+t_{m}^{\mathrm{c}}=T,$
such that the constraints of the problem \eqref{eq:p1-3} are still
satisfied.
\end{IEEEproof}
Proposition 2 indicates that the time resource shall be fully utilized,
i.e., the time allocated for computing the offloaded data shall last
until the end of the considered time. From Proposition 1 and Proposition
2, the feasible solution to the problem \eqref{eq:p1-3} can satisfy
\begin{align}
 & f_{k}=\frac{r_{k}T}{T-t_{m}^{\mathrm{o}}},\forall k\in\mathcal{K}_{m},\label{eq:fmk}\\
 & t_{m}^{\mathrm{c}}=T-t_{m}^{\mathrm{o}}.\label{eq:tmc}
\end{align}
By inserting \eqref{eq:fmk} and \eqref{eq:tmc} into the problem
\eqref{eq:p1-3}, we get 
\begin{subequations}
\label{eq:p1-8}
\begin{alignat}{1}
\mathrm{Find} & \:t_{m}^{\mathrm{o}},\mathbf{p}_{m},\boldsymbol{\pi}_{m}\label{eq:p1-2-o-1}\\
\mathrm{s.t.} & \:t_{m}^{\mathrm{o}}\geq\max_{k\in\mathcal{K}_{m}}\frac{T\eta}{\sum_{n\in\mathcal{N}}\sum_{i=1}^{2}\beta_{n,k}r_{m,n,k,i}},\\
 & \:t_{m}^{\mathrm{o}}\leq\frac{TF_{m}}{F_{m}+\sum_{k\in\mathcal{K}_{m}}\sum_{n\in\mathcal{N}}\sum_{i=1}^{2}\beta_{n,k}r_{m,n,k,i}},\\
 & \:\eqref{eq:p1-3-c1},\eqref{eq:p1-3-c8},\eqref{eq:p1-3-c9}.\nonumber 
\end{alignat}
\end{subequations}
Since we only need to find a feasible solution to the problem \eqref{eq:p1-8},
a feasible $t_{m}^{\mathrm{o}}$ can take any value within the feasible
region, and we choose its maximum allowable value as its feasible
value, i.e.
\begin{equation}
t_{m}^{\mathrm{o}}=\frac{TF_{m}}{F_{m}+\sum_{k\in\mathcal{K}_{m}}\sum_{n\in\mathcal{N}}\sum_{i=1}^{2}\beta_{n,k}r_{m,n,k,i}}.\label{eq:tmo}
\end{equation}
The expression of $t_{m}^{\mathrm{o}}$ in \eqref{eq:tmo} implies
that the time required for data offloading increases as $F_{m}$ increases.
The reason for this is that a higher total computing frequency can
reduce the time needed for computing at the MEC server and leave more
time for data offloading. The expression in \eqref{eq:tmo} also implies
that the data offloading time is a decreasing function of the computation
offloading rate. This is expected since a higher computation offloading
rate can let the device offload the data more quickly with a given
amount of offloaded data.

By substituting \eqref{eq:tmo} into \eqref{eq:fmk} and \eqref{eq:tmc},
we get
\begin{align}
f_{k} & =\frac{F_{m}\sum_{n\in\mathcal{N}}\sum_{i=1}^{2}\beta_{n,k}r_{m,n,k,i}}{\sum_{k\in\mathcal{K}_{m}}\sum_{n\in\mathcal{N}}\sum_{i=1}^{2}\beta_{n,k}r_{m,n,k,i}},\forall k\in\mathcal{K}_{m},\label{eq:fmk-2}\\
t_{m}^{\mathrm{c}} & =\frac{T\sum_{k\in\mathcal{K}_{m}}\sum_{n\in\mathcal{N}}\sum_{i=1}^{2}\beta_{n,k}r_{m,n,k,i}}{F_{m}+\sum_{k\in\mathcal{K}_{m}}\sum_{n\in\mathcal{N}}\sum_{i=1}^{2}\beta_{n,k}r_{m,n,k,i}}.\label{eq:tmc-2}
\end{align}
The above expression in \eqref{eq:fmk-2} indicates that the computing
frequency allocated to the device is proportional to its computation
offloading rate. This is due to the fact that a higher computation
offloading rate can let the device offload more data to the MEC server,
and this will require a higher computing frequency at the MEC server
to compute the offloaded data. The above expression in \eqref{eq:tmc-2}
suggests that the time required for data computing increases as the
computation offloading rate increases. This is because a higher computation
offloading rate can let the device offload more data to the MEC server.

By inserting \eqref{eq:tmo} into the problem \eqref{eq:p1-8}, we
have
\begin{subequations}
\label{eq:p1-4}
\begin{alignat}{1}
\mathrm{Find} & \:\mathbf{p}_{m},\boldsymbol{\pi}_{m}\\
\mathrm{s.t.} & \:F_{m}\sum_{n\in\mathcal{N}}\sum_{i=1}^{2}\beta_{n,k}r_{m,n,k,i}-\eta\sum_{k^{'}\in\mathcal{K}_{m}}\sum_{n\in\mathcal{N}}\sum_{i=1}^{2}\beta_{n,k^{'}}\nonumber \\
 & \:\times r_{m,n,k^{'},i}\geq\eta F_{m},\forall k\in\mathcal{K}_{m},\label{eq:p1-4-c1}\\
 & \:\eqref{eq:p1-3-c1},\eqref{eq:p1-3-c8},\eqref{eq:p1-3-c9}.\nonumber 
\end{alignat}
\end{subequations}

The problem \eqref{eq:p1-4} is a non-convex mixed integer problem.
To optimally solve the problem \eqref{eq:p1-4}, an exhaustive search
of $\boldsymbol{\pi}_{m}$ shall be performed, where in each search
the problem \eqref{eq:p1-4} under the given $\boldsymbol{\pi}_{m}$
is solved. Since there are $(2|\mathcal{K}_{m}|)!$ possible decoding
orders for $2|\mathcal{K}_{m}|$ sub-messages, the exhaustive search
is of high complexity even for a very small set $\mathcal{K}_{m}$
. Thus, we propose a low-complexity solution by optimizing $\mathbf{p}_{m}$
and $\boldsymbol{\pi}_{m}$ based on the alternating optimization
method. Specifically, we alternatively optimize $\mathbf{p}_{m}$
with given $\boldsymbol{\pi}_{m}$, and optimize $\boldsymbol{\pi}_{m}$
with given $\mathbf{p}_{m}$, until a feasible solution to the problem
\eqref{eq:p1-4} is found, or the optimization converges, or the maximum
number of iterations is reached.

For the problem \eqref{eq:p1-4} of optimizing $\mathbf{p}_{m}$ with
given $\boldsymbol{\pi}_{m}$, it is a non-convex problem. For the
convenience of later discussion, $r_{m,n,k,i}$ is rewritten as
\begin{equation}
r_{m,n,k,i}=w_{m,n,k,i}-v_{m,n,k,i},
\end{equation}
where
\begin{align}
w_{m,n,k,i}= & B\log_{2}\left(\sigma^{2}B+\sum_{(k^{'},i^{'})\in\mathcal{S}_{m,n,k,i}}h_{m,n,k^{'}}p_{k^{'},i^{'}}\right.\nonumber \\
 & \left.\vphantom{\sum_{(k^{'},i^{'})\in\mathcal{S}_{m,n,k,i}}}+h_{m,n,k}p_{k,i}\right)\\
v_{m,n,k,i}= & B\log_{2}\left(\sigma^{2}B+\sum_{(k^{'},i^{'})\in\mathcal{S}_{m,n,k,i}}h_{m,n,k^{'}}p_{k^{'},i^{'}}\right).
\end{align}

\noindent It can be easily verified that both $w_{m,n,k,i}$ and $v_{m,n,k,i}$
are concave with respect to $\mathbf{p}_{m}$. Based on $w_{m,n,k,i}$
and $v_{m,n,k,i}$, the constraint \eqref{eq:p1-4-c1} in the problem
\eqref{eq:p1-4} can be rewritten as
\begin{equation}
z_{m,k}-l_{m,k}\geq\eta F_{m},\forall k\in\mathcal{K}_{m},\label{eq:p1-4-c2}
\end{equation}
where 
\begin{align}
z_{m,k}= & (F_{m}-\eta)\sum_{n\in\mathcal{N}}\sum_{i=1}^{2}\beta_{n,k}w_{m,n,k,i}\nonumber \\
 & +\eta\sum_{k^{'}\in\mathcal{K}_{m},k^{'}\neq k}\sum_{n\in\mathcal{N}}\sum_{i=1}^{2}\beta_{n,k^{'}}v_{m,n,k^{'},i},\\
l_{m,k}= & (F_{m}-\eta)\sum_{n\in\mathcal{N}}\sum_{i=1}^{2}\beta_{n,k}v_{m,n,k,i}\nonumber \\
 & +\eta\sum_{k^{'}\in\mathcal{K}_{m},k^{'}\neq k}\sum_{n\in\mathcal{N}}\sum_{i=1}^{2}\beta_{n,k^{'}}w_{m,n,k^{'},i}.\label{eq:lmk}
\end{align}
It can be shown that $F_{m}-\eta>0,$ otherwise we will have $z_{m,k}-l_{m,k}<0$
as can be seen from \eqref{eq:p1-4-c1}. Thus, both $z_{m,k}$ and
$l_{m,k}$ are concave functions of $\mathbf{p}_{m}$. Moreover, it
can be easily shown that the constraint \eqref{eq:p1-4-c2} is equivalent
to
\begin{equation}
\min_{k\in\mathcal{K}_{m}}z_{m,k}-l_{m,k}\geq\eta F_{m}.\label{eq:p1-4-c3}
\end{equation}
Thus, the problem \eqref{eq:p1-4} of optimizing $\mathbf{p}_{m}$
with given $\boldsymbol{\pi}_{m}$ is re-expressed as
\begin{alignat}{1}
\mathrm{Find} & \:\mathbf{p}_{m}\label{eq:p1-4-1}\\
\mathrm{s.t.} & \:\eqref{eq:p1-3-c1},\eqref{eq:p1-3-c8},\eqref{eq:p1-4-c3}.\nonumber 
\end{alignat}
Since the problem \eqref{eq:p1-4-1} is feasible if and only if the
constraint \eqref{eq:p1-4-c3} can be satisfied, the problem \eqref{eq:p1-4-1}
can be reformulated as the following problem of maximizing the left-hand-side
of the constraint \eqref{eq:p1-4-c3} as given by 
\begin{alignat}{1}
\mathrm{\max_{\mathbf{p}_{m}}} & \:\min_{k\in\mathcal{K}_{m}}z_{m,k}-l_{m,k}\label{eq:p1-5}\\
\mathrm{s.t.} & \:\eqref{eq:p1-3-c1},\eqref{eq:p1-3-c8},\nonumber 
\end{alignat}

\noindent Note that the problem \eqref{eq:p1-4-1} is feasible if
and only if the optimal objective function value in \eqref{eq:p1-5}
is not smaller than $\eta F_{m}.$ Since the objective function in
\eqref{eq:p1-5} is in the form of the difference of two concave functions
and the constraints are linear, the problem \eqref{eq:p1-5} is non-convex
and its optimal solution is still hard to find. Hence, we resort to
the SCA method to solve the problem \eqref{eq:p1-5} suboptimally
in an iterative manner. The main idea of the SCA method is to iteratively
approximate the non-convex problem as a convex problem and solve the
approximate convex problem until the solution converges. For the problem
\eqref{eq:p1-5}, since its objective function is in the form of the
difference of two concave functions, we can approximate the objective
function as a concave function by approximating the latter concave
function with a linear function based on the first-order Taylor series
expansion. Specifically, let $\mathbf{p}_{m}^{(t)}$ denote the solution
in the $t$-th iteration. Then, given $\mathbf{p}_{m}^{(t-1)}$ derived
in the $(t-1)$-th iteration, the $v_{m,n,k,i}$ in $l_{m,k}$ is
approximated using the first-order Taylor series expansion as
\begin{align}
 & \hat{v}_{m,n,k,i}=v_{m,n,k,i}^{(t-1)}+\frac{\partial v_{m,n,k,i}^{(t-1)}}{\partial\mathbf{p}_{m}}(\mathbf{p}_{m}-\mathbf{p}_{m}^{(t-1)}),\label{eq:approxV}
\end{align}
where $v_{m,n,k,i}^{(t-1)}$ and $\frac{\partial v_{m,n,k,i}^{(t-1)}}{\partial\mathbf{p}_{m}}$
are the values of $v_{m,n,k,i}$ and $\frac{\partial v_{m,n,k,i}}{\partial\mathbf{p}_{m}}$
with $\mathbf{p}_{m}=\mathbf{p}_{m}^{(t-1)},$ respectively. The $\frac{\partial v_{m,n,k,i}}{\partial\mathbf{p}_{m}}=\left\{ \frac{\partial v_{m,n,k,i}}{\partial p_{k^{'},i^{'}}},\forall k^{'}\in\mathcal{K}_{m},\forall i^{'}\right\} $
can be obtained as \eqref{eq:derivative-v}. 
\begin{figure*}[!t]
\noindent 
\begin{align}
\frac{\partial v_{m,n,k,i}}{\partial p_{k^{'},i^{'}}}= & \begin{cases}
\frac{Bh_{m,n,k^{'}}}{\left(\sigma^{2}B+\sum_{(k^{''},i^{''})\in\mathcal{S}_{m,n,k,i}}h_{m,n,k^{''}}p_{k^{''},i^{''}}\right)\ln2}, & (k^{'},i^{'})\in\mathcal{S}_{m,n,k,i},\\
0, & (k^{'},i^{'})\notin\mathcal{S}_{m,n,k,i}.
\end{cases}\label{eq:derivative-v}
\end{align}

\noindent \rule[0.5ex]{1\textwidth}{0.5pt}
\end{figure*}
Meanwhile, $w_{m,n,k^{'},i},\forall k^{'}\in\mathcal{K}_{m},k^{'}\neq k$
in $l_{m,k}$ is also approximated using the first-order Taylor series
expansion as
\begin{align}
 & \hat{w}_{m,n,k^{'},i}=w_{m,n,k^{'},i}^{(t-1)}+\frac{\partial w_{m,n,k^{'},i}^{(t-1)}}{\partial\mathbf{p}_{m}}(\mathbf{p}_{m}-\mathbf{p}_{m}^{(t-1)}),\label{eq:approxW}
\end{align}
where $w_{m,n,k^{'},i}^{(t-1)}$ and $\frac{\partial w_{m,n,k^{'},i}^{(t-1)}}{\partial\mathbf{p}_{m}}$
are the values of $w_{m,n,k^{'},i}$ and $\frac{\partial w_{m,n,k^{'},i}}{\partial\mathbf{p}_{m}}$
with $\mathbf{p}_{m}=\mathbf{p}_{m}^{(t-1)},$ respectively. The $\frac{\partial w_{m,n,k^{'},i}}{\partial\mathbf{p}_{m}}=\left\{ \frac{\partial w_{m,n,k^{'},i}}{\partial p_{k^{''},i^{''}}},\forall k^{''}\in\mathcal{K}_{m},i^{''}\right\} $
can be obtained as given by \eqref{eq:derivative-w}. 
\begin{figure*}[!t]
\noindent 
\begin{equation}
\frac{\partial w_{m,n,k^{'},i}}{\partial p_{k^{''},i^{''}}}=\begin{cases}
\frac{Bh_{m,n,k^{''}}}{\left(\sigma^{2}B+\sum_{(k^{'''},i^{'''})\in\mathcal{S}_{m,n,k^{'},i}}h_{m,n,k^{'''}}p_{k^{'''},i^{'''}}+h_{m,n,k^{'}}p_{k^{'},i}\right)\ln2}, & (k^{''},i^{''})\in\mathcal{S}_{m,n,k^{'},i}\cup\{k^{'},i\},\\
0, & (k^{''},i^{''})\notin\mathcal{S}_{m,n,k^{'},i}\cup\{k^{'},i\}.
\end{cases}\label{eq:derivative-w}
\end{equation}

\noindent \rule[0.5ex]{1\textwidth}{0.5pt}
\end{figure*}
Thus, the problem \eqref{eq:p1-5} in the $(t-1)$-th iteration is
approximated as
\begin{subequations}
\label{eq:p1-6}
\begin{alignat}{1}
\mathrm{\max} & \:\min_{k\in\mathcal{K}_{m}}z_{m,k}-\hat{l}_{m,k}\label{eq:p1-6-o}\\
\mathrm{s.t.} & \:\eqref{eq:p1-3-c1},\eqref{eq:p1-3-c8},\nonumber \\
\mathrm{o.v.} & \:\mathbf{p}_{m},
\end{alignat}
\end{subequations}
where $\hat{l}_{m,k}$ is given by \eqref{eq:lmk} with $v_{m,n,k,i}$
replaced by $\hat{v}_{m,n,k,i}$ in \eqref{eq:approxV} and $w_{m,n,k^{'},i}$
replaced with $\hat{w}_{m,n,k^{'},i}$ in \eqref{eq:approxW}. Since
both $w_{m,n,k,i}$ and $v_{m,n,k,i}$ are concave functions, $z_{m,k}-\hat{l}_{m,k}$
is concave, and its pointwise minimum function $\min_{k\in\mathcal{K}_{m}}z_{m,k}-\hat{l}_{m,k}$
is also concave \cite{convexop2004}. Thus, the problem \eqref{eq:p1-6}
is a convex problem and can be solved efficiently by the interior
point method using CVX \cite{cvx}.

For the problem \eqref{eq:p1-4} of optimizing $\boldsymbol{\pi}_{m}$
with given $\mathbf{p}_{m}$, the optimal solution can be still obtained
via the exhaustive searching of $\boldsymbol{\pi}_{m},$ which is
of high complexity. Thus, we propose a particular SIC decoding order
$\boldsymbol{\pi}_{m}$ for low-complexity realization. Specifically,
the SIC decoding order of the sub-messages $s_{k,i},\forall k\in\mathcal{K}_{m},\forall i\in\{0,1\}$
is proposed to be in the descending order of $h_{m,n,k}p_{k,i}$,
i.e., $\pi_{k,i}<\pi_{k^{'},i^{'}}$ for $h_{m,n,k}p_{k,i}>h_{m,n,k^{'}}p_{k^{'},i^{'}}$.
The intuition of choosing such a SIC decoding order is that a sub-message
with higher $h_{m,n,k}p_{k,i}$ causes higher residual interference
to the sub-messages whose SIC decoding order is larger than this sub-messages,
and thus decoding the sub-message with higher $h_{m,n,k}p_{k,i}$
first will reduce the residual interference caused to other sub-messages
and can result in a higher MCOR.

For initializing the alternating optimization of $\mathbf{p}_{m}$
and $\boldsymbol{\pi}_{m}$, we further propose a heuristic decoding
order $\boldsymbol{\pi}_{m}$ that is independent of $\mathbf{p}_{m}$.
Specifically, we propose to first decode all the sub-messages $s_{k,1},\forall k\in\mathcal{K}_{m}$
and then decode all the sub-messages $s_{k,2},\forall k\in\mathcal{K}_{m}$,
i.e., $\pi_{k,1}<\pi_{k^{'},2},\forall k\in\mathcal{K}_{m},k^{'}\in\mathcal{K}_{m}$,
where the SIC decoding order of the sub-messages $s_{k,i},\forall k\in\mathcal{K}_{m}$
for each $i\in\{1,2\}$ is proposed to be in the descending order
of $h_{m,n,k}$, i.e., $\pi_{k,i}<\pi_{k^{'},i^{'}},\forall k\in\mathcal{K}_{m},k^{'}\in\mathcal{K}_{m},k\neq k^{'}$
for $h_{m,n,k}>h_{m,n,k^{'}}$. The intuition of choosing such a SIC
decoding order is that a sub-message with higher $h_{m,n,k}$ will
more likely cause higher residual interference to the sub-messages
whose SIC decoding order is larger than this sub-messages, and thus
decoding the sub-message with higher $h_{m,n,k}$ first will highly
likely reduce the residual interference caused to other sub-messages.
Meanwhile, decoding all sub-messages $s_{k,1},\forall k$ prior to
all sub-messages $s_{k,2},\forall k$ can effectively adjust the rate
splitting between two sub-messages of each device, since it was shown
that it is optimal to separate the two sub-messages in the case of
two devices \cite{yang2022sum}. The heuristic SIC decoding order
also can let the device with inferior channel conditions suffer less
interference from other devices, thus improves the fairness among
devices and leads to a higher MCOR.

The proposed algorithm for optimizing $\mathbf{t}^{\mathrm{o}},$
$\mathbf{t}^{\mathrm{c}},$ $\mathbf{f},$ $\mathbf{p}$ and $\boldsymbol{\pi}$
is summarized in Algorithm \ref{alg:1}. Note that although the SCA
method has been used in existing works \cite{yang2021optimization,zhou2022rate,van2023evolutionary,abbasi2023transmission},
we use the SCA method to optimize the power allocation in uplink RSMA
considering the max-min fairness among multiple devices, whereas \cite{yang2021optimization,zhou2022rate,van2023evolutionary}
optimized the resource allocation in downlink RSMA and \cite{abbasi2023transmission}
optimized the power allocation of only two devices in uplink RSMA
without considering the fairness among them.

\begin{algorithm}[!t]
\begin{algorithmic}[1]
\STATE Initialize $\eta_{\mathrm{min}}$ and  $\eta_{\mathrm{max}}$.
\REPEAT
\STATE $\eta=\frac{\eta_{\mathrm{min}}+\eta_{\mathrm{max}}}{2}$.
\FORALL{$m$ such that $m\in\mathcal{M}$}

\STATE Initialize $\boldsymbol{\pi}_m$ as $\pi_{k,1}<\pi_{k^{'},2},\forall k\in\mathcal{K}_{m},k^{'}\in\mathcal{K}_{m}$ and $\pi_{k,i}<\pi_{k^{'},i},\forall k\in\mathcal{K}_{m},k^{'}\in\mathcal{K}_{m},k\neq k^{'}$ for $h_{m,n,k}>h_{m,n,k^{'}}$.

\REPEAT

\STATE Initialize $\mathbf{p}_{m}^{(0)}$ and $t=0$.

\REPEAT

\STATE Derive $\mathbf{p}_{m}^{(t+1)}$ by solving the problem \eqref{eq:p1-6} with given $\mathbf{p}_{m}^{(t)}$ using CVX.
\STATE $t=t+1$.
\UNTIL{the difference between $\mathbf{p}_{m}^{(t+1)}$ and $\mathbf{p}_{m}^{(t)}$ is within the desired accuracy.}

\STATE Update $\boldsymbol{\pi}_m$ as $\pi_{k,i}<\pi_{k^{'},i^{'}},\forall k\in\mathcal{K}_{m},k^{'}\in\mathcal{K}_{m},i\in\{1,2\},i^{'}\in\{1,2\}$ for $h_{m,n,k}p_{k,i}>h_{m,n,k^{'}}p_{k^{'},i^{'}}$.
\UNTIL{the objective function value in \eqref{eq:p1-5} is not smaller than $\eta F_{m}$ or converges, or the maximum number of iterations is reached.}

\STATE Obtain $t_{m,o}$, $f_{k},\forall k\in\mathcal{K}_{m}$ and $t_{m,c}$   from \eqref{eq:tmo}, \eqref{eq:fmk-2} and \eqref{eq:tmc-2}, respectively.

\ENDFOR

\IF{the objective function value in \eqref{eq:p1-5} is not smaller than $\eta F_{m}$ for all $m\in\mathcal{M}$}
\STATE $\eta_{\mathrm{min}}=\eta$.
\ELSE
\STATE $\eta_{\mathrm{max}}=\eta$.
\ENDIF

\UNTIL{$\eta$ converges.}
\STATE Output: $\mathbf{t}^{\mathrm{o}},$ $\mathbf{t}^{\mathrm{c}},$ $\mathbf{f},$ $\mathbf{p}$ and $\boldsymbol{\pi}$.
\end{algorithmic}\caption{Proposed algorithm based on the alternating optimization method, the
bisection search method and the SCA method for optimizing $\mathbf{t}^{\mathrm{o}},$
$\mathbf{t}^{\mathrm{c}},$ $\mathbf{f},$ $\mathbf{p}$ and $\boldsymbol{\pi}$
given $\boldsymbol{\alpha}$ and $\boldsymbol{\beta}$.\label{alg:1}}
\end{algorithm}

\textbf{\textcolor{black}{Convergence analysis}}: It can be shown
that the bisection search method in the outermost loop from Step 2
to Step 21 of Algorithm \ref{alg:1} clearly converges and its convergence
does not depend on the objective function value in \eqref{eq:p1-5}
\cite{convexop2004}. This means that the convergence of the alternating
optimization method in the middle loop from Step 6 to Step 13 and
the convergence of the SCA method in the innermost loop from Step
8 to Step 11 of Algorithm \ref{alg:1} does not affect the convergence
of the bisection search method. In addition, the convergence of the
alternating optimization method in the middle loop from Step 6 to
Step 13 is guaranteed by setting the maximum number of iterations.
This means that the convergence of Algorithm \ref{alg:1} depends
on the convergence of the SCA method in the innermost loop from Step
8 to Step 11. Since we obtain the optimal solution to the problem
\eqref{eq:p1-6} in Step 9, the objective function value in \eqref{eq:p1-6}
is non-decreasing as the iteration continues in the SCA method. Besides,
the objective function value in \eqref{eq:p1-6} is upper-bounded.
Thus, the SCA method converges to a local optimal solution, which
means that Algorithm \ref{alg:1} also converges.

\textbf{\textcolor{black}{Complexity analysis}}: The complexity of
the bisection search method for obtaining $\ensuremath{\eta}$ is
$\mathcal{O}\left(\log_{2}\frac{\bar{\eta}}{\varepsilon}\right)$
\cite{convexop2004}, where $\varepsilon$ is the error tolerance
and $\bar{\eta}$ is a value that is large enough to serve as an upper-bound
for the optimal $\eta$. Since the convex problem \eqref{eq:p1-6}
has $2|\mathcal{K}_{m}|$ optimization variables and $|\mathcal{K}_{m}|$
inequality constraints, the complexity of solving the convex problem
\eqref{eq:p1-6} is $\mathcal{O}\left(|\mathcal{K}_{m}|^{3}\right)$
\cite{convexop2004,yang2022sum}. Let $D_{1}$ and $D_{2}$ denote
the maximum number of iterations of the alternating optimization method
and the number of iterations for the SCA method to converge, respectively,
which are both independent of the number of devices. Therefore, the
total complexity of the proposed algorithm in Algorithm \ref{alg:1}
is approximately $\mathcal{O}(D_{1}D_{2}\sum_{m\in\mathcal{M}}|\mathcal{K}_{m}|^{3}\log_{2}\frac{\bar{\eta}}{\varepsilon})$.

\section{Channel and MEC Server Allocation Using Matching Theory\label{sec:MEC-Allocation-and}}

\begin{figure}[!t]
\centering\includegraphics[width=0.8\columnwidth]{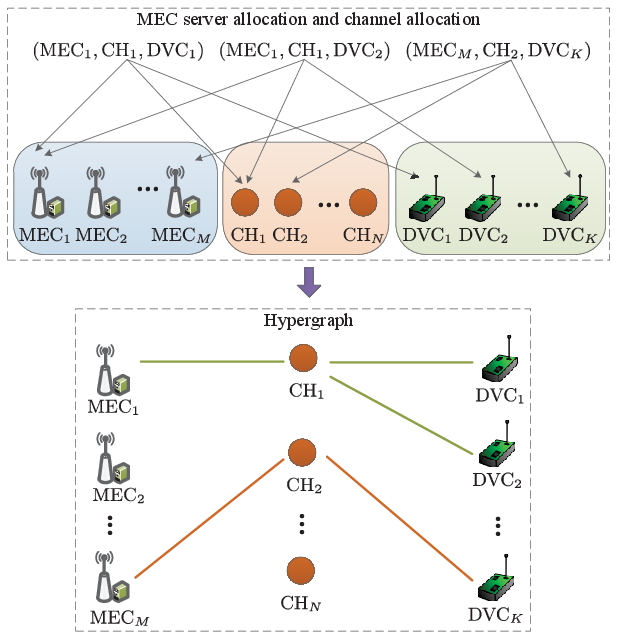}

\caption{Graphical expression of MEC server allocation and channel allocation.\label{fig:Graphical-expression-of}}
\end{figure}

This section investigates the channel and MEC server allocation problem
in the problem \eqref{eq:p1} with given $\mathbf{t}^{\mathrm{o}},$
$\mathbf{t}^{\mathrm{c}},$ $\mathbf{f},$ $\mathbf{p}$ and $\boldsymbol{\pi}$.
The problem is given by
\begin{alignat}{1}
\max_{\boldsymbol{\alpha},\boldsymbol{\beta}} & \:\min_{k\in\mathcal{K}}r_{k}\label{eq:p2}\\
\mathrm{s.t.} & \:\eqref{eq:p1-c0},\eqref{eq:p1-c1},\eqref{eq:p1-c1-1},\eqref{eq:p1-c7},\eqref{eq:p1-c8},\nonumber 
\end{alignat}
The problem \eqref{eq:p2} is a nonlinear integer programming problem,
whose optimal solution can be obtained by the exhaustive search method.
However, the complexity of the optimal solution is unbearable. Thus,
a matching theory based algorithm is proposed to obtain a suboptimal
solution. From the graphical point of view, the relationship among
MEC servers, channels and devices is represented in the top part of
Fig. \ref{fig:Graphical-expression-of}. For the convenience of later
analysis, a hypergraph based representation is shown at the bottom
part of Fig. \ref{fig:Graphical-expression-of}, where lines in different
colors represent different hyperedges. Generally speaking, hypergraph
is a generalized graph, where a hyperedge consists of any subset of
the vertices, whereas in a traditional graph, an edge connects two
vertices. Specifically, in the considered hypergraph in Fig. \ref{fig:Graphical-expression-of},
MEC servers, channels and devices are vertices, and a hyperedge consists
of an MEC server, a channel and multiple devices, which means that
the MEC server and the channel are allocated to the devices. Hence,
the problem of MEC server allocation and channel allocation is transformed
into a hypergraph matching problem \cite{zhang2017hypergraph}. 

As illustrated in Fig. \ref{fig:Graphical-expression-of}, when $\mathrm{MEC}_{m}$,
$\mathrm{CH}_{n}$ and $\mathrm{DVC}_{k},k\in\mathcal{K}_{m,n}$ are
matched with each other, a hyperedge is formed, denoted by a matching
triple $(\mathrm{MEC}_{m},\mathrm{CH}_{n},(\mathrm{DVC}_{k},k\in\mathcal{K}_{m,n}))\subset\mathcal{M}\cup\mathcal{N}\cup\mathcal{K}$,
where $\mathcal{K}_{m,n}$ is the set of devices allocated with MEC
server $\mathrm{MEC}_{m}$ and channel $\mathrm{CH}_{n}$. Since finding
the optimal solution to the hypergraph matching problem is intractable,
we propose a low-complexity algorithm to achieve a suboptimal solution
by decomposing the problem into two simpler problems, i.e., the channel
and device matching problem and the (channel, device) and MEC server
matching problem. Specifically, allocating a channel to each device
is considered in the channel and device matching problem, which is
formulated as a one-to-many matching problem between channels and
devices, i.e., one channel can be allocated to multiple devices. Then,
allocating MEC servers to (channel, device) units is considered in
the (channel, device) and MEC server matching problem, which is also
formulated as a one-to-many matching problem. In what follows, we
adopt matching theory to solve the two matching problems.

\subsection{Preliminaries for Matching Theory}

In a traditional two-sided matching, there are two finite and disjoint
sets denoted by $\mathcal{X}=\{x_{1},\ldots,x_{A}\}$ and $\mathcal{Y}=\{y_{1},\ldots,y_{B}\},$
respectively, where each element in the two sets has a preference
list with respect to another set. In this paper, the preference can
be the MCOR, which represents the priorities of its selection among
different candidates. Specifically, we denote $y_{1}\succ_{x_{a}}y_{2}$
if $x_{a}$ prefers $y_{1}$ to $y_{2}$. Due to the randomness of
channel gains, any two candidates cannot achieve the same MCOR and
thus they can be compared to get a preferred one, which means that
the complete ordering property is satisfied by the preference lists.
In addition, it is easily shown that if $y_{1}\succ_{x_{a}}y_{2}$
and $y_{2}\succ_{x_{a}}y_{3}$, then $y_{1}\succ_{x_{a}}y_{3}$, which
means that the transitive property is also satisfied by the preference
list. Based on the above discussions, the following definition is
presented.

\textbf{\textcolor{black}{Definition 1}}: Let $\xi$ denote the one-to-many
mapping between the set $\mathcal{X}$ and set $\mathcal{Y}$. A mapping
pair of $x_{a}\in\mathcal{X}$ and $y_{b}\in\mathcal{Y}$ is denoted
as:
\begin{itemize}
\item $\xi_{a}\subseteq\mathcal{Y};$
\item $\xi_{b}\subseteq\mathcal{X};$
\item $|\xi_{a}|\leq C;$
\item $|\xi_{b}|\leq1;$
\item $y_{b}\in\xi_{a}\Leftrightarrow x_{a}\in\xi_{b}$,
\end{itemize}
where the first and the second conditions mean that the elements matched
with $x_{a}$ belong to set $\mathcal{Y}$, and the elements matched
with $y_{b}$ belong to set $\mathcal{X}$, respectively, the third
and the fourth conditions define the maximum number of elements that
can be matched with $x_{a}$ and $y_{b}$, respectively, and the final
condition means that if $y_{b}$ is matched with $x_{a},$ then $x_{a}$
is also matched with $y_{b}$. If the matching has externalities or
peer effects where the gain from a matching pair depends on the matching
of other players, the matching decisions will be dynamically affected
by each other, and it is not straightforward to achieve a stable matching.
Therefore, the concept of swap matching is introduced to achieve the
exchange stability of matching. Specifically, a swap matching is defined
as $\xi_{a}^{a^{'}}=\{\xi\backslash\{(x_{a},y_{b^{'}}),(x_{a^{'}},y_{b})\}\cup\{(x_{a},y_{b}),(x_{a^{'}},y_{b^{'}})\}\},$
where $\xi_{a}=y_{b^{'}}$ and $\xi_{a^{'}}=y_{b}$, and the two-sided
exchange stability is defined as follows.

\textbf{\textcolor{black}{Definition 2}}: A matching $\xi$ is two-sided
exchange stable if and only if there does not exist a swap blocking
pair $(x_{a},x_{a^{'}})$ with $y_{b^{'}}=\xi_{a}$ and $y_{b}=\xi_{a^{'}}$:
\begin{itemize}
\item $\forall X\in\{x_{a},x_{a^{'}},y_{b^{'}},y_{b}\}$, the utility under
matching $\xi_{a}^{a^{'}}$ is not smaller than that under matching
$\xi$;
\item $\exists X\in\{x_{a},x_{a^{'}},y_{b^{'}},y_{b}\}$, the utility under
matching $\xi_{a}^{a^{'}}$ is larger than that under matching $\xi$,
\end{itemize}

According to the above definition, a swap blocking pair means that
the utility of players after swapping will not decrease, and at least
one player's utility will increase. Therefore, exchange stability
is achieved if no player can benefit from swapping without sacrificing
other players' benefits.

\subsection{Channel Allocation Problem\label{subsec:MEC-Server-Allocation}}

Here, the channel allocation problem is reconstructed as a one-to-many
matching problem. From \eqref{eq:rate-k}, since the computation offloading
rate of each device is related to the set of devices on the same channel
associated with the same MEC server, each device considers not only
which channel to match with but also other devices' decisions. Thus,
the channel allocation problem is a one-to-many matching problem with
externalities. Specifically, the preference of each device $\mathrm{DVC}_{k}$
over channel $\mathrm{CH}_{n}$ with a set of devices $\mathcal{K}_{n}$
already allocated with channel $\mathrm{CH}_{n}$ is chosen as the
MCOR over all MEC servers, which can be expressed as
\begin{equation}
\mathcal{P}_{k,n,\mathcal{K}_{n}}=\min_{m\in\mathcal{M}}\frac{t_{m}^{\mathrm{o}}}{T}\sum_{i=1}^{2}r_{m,n,k,i},\label{eq:P-device}
\end{equation}
where $\mathcal{S}_{m,n,k,i}$ in $r_{m,n,k,i}$ is re-expressed as
\begin{equation}
\mathcal{S}_{m,n,k,i}=\{(k^{'},i^{'})|k^{'}\in\mathcal{K}_{n},\pi_{k^{'},i^{'}}>\pi_{k,i},\forall k^{'},i^{'}\}.\label{eq:Ski-2}
\end{equation}
Then, the preference of each channel $\mathrm{CH}_{n}$ over a set
of devices $\mathcal{K}_{n}^{'}$ with a set of devices $\mathcal{K}_{n}$
already allocated with it is expressed as
\begin{equation}
\mathcal{P}_{\mathcal{K}_{m}^{'},n,\mathcal{K}_{n}}=\min_{k\in\mathcal{K}_{n}^{'}}\mathcal{P}_{k,n,\mathcal{K}_{n}\cup\mathcal{K}_{n}^{'}}.\label{eq:P-MEC}
\end{equation}
Based on the above defined preferences, for a given $\mathrm{DVC}_{k}$,
any two channels $\mathrm{CH}_{n}$ and $\mathrm{CH}_{n^{'}}$, any
two matchings $\xi$ and $\xi^{'}$, we have
\begin{equation}
(\mathrm{CH}_{n},\xi)\succ_{\mathrm{DVC}_{k}}(\mathrm{CH}_{n^{'}},\xi^{'})\Leftrightarrow\mathcal{P}_{k,n,\mathcal{K}_{n}}(\xi)>\mathcal{P}_{k,n^{'},\mathcal{K}_{n^{'}}}(\xi^{'}),
\end{equation}
where $\mathcal{P}_{k,n,\mathcal{K}_{n}}(\xi)$ is the preference
of $\mathrm{DVC}_{k}$ over $\mathrm{CH}_{n}$ with a set of already
associated devices $\mathcal{K}_{n}$ under the matching $\xi.$ The
above expression indicates that $\mathrm{DVC}_{k}$ prefers $\mathrm{CH}_{n}$
in $\xi$ to $\mathrm{CH}_{n^{'}}$ in $\xi^{'}$, provided that the
preference $\mathcal{P}_{k,n,\mathcal{K}_{n}}(\xi)$ is larger than
the preference $\mathcal{P}_{k,n^{'},\mathcal{K}_{n^{'}}}(\xi^{'})$.
Similarly, for a given $\mathrm{CH}_{n}$, any two sets of devices
$\mathcal{K}_{1}$ and $\mathcal{K}_{2}$, any two matchings $\xi$
and $\xi^{'}$ with $\mathcal{K}_{1}=\xi_{n}$ and $\mathcal{K}_{2}=\xi_{n}^{'}$,
we have
\begin{equation}
(\mathcal{K}_{1},\xi)\!\succ_{\mathrm{CH}_{n}}(\mathcal{K}_{2},\xi^{'})\!\Leftrightarrow\!\mathcal{P}_{\mathcal{K}_{1},n,\mathcal{K}_{n}}(\xi)\!>\!\mathcal{P}_{\mathcal{K}_{2},n,\mathcal{K}_{n}}(\xi^{'}),
\end{equation}
where $\mathcal{P}_{\mathcal{K}_{i},n,\mathcal{K}_{n}}(\xi)$ is the
preference of $\mathrm{CH}_{n}$ over a set of devices $\mathcal{K}_{i}$
with a set of already associated devices $\mathcal{K}_{n}$ under
the matching $\xi$. This means that $\mathrm{CH}_{n}$ prefers the
set of devices $\mathcal{K}_{1}$ to $\mathcal{K}_{2}$ only when
higher MCOR can be achieved on $\mathrm{CH}_{n}$ from $\mathcal{K}_{1}$
than that from $\mathcal{K}_{2}$.

For performing channel allocation, the Gale-Shapley algorithm \cite{gale1962college}
is utilized to construct an initial matching, as shown in Algorithm
\ref{alg2}. Specifically, the algorithm consists of the request and
the response phases. In the request phase, all the channels simultaneously
request the most preferred device that they were not rejected by.
Then, in the response phase, each device that is requested by more
than one channel chooses the most preferred one and rejects all the
others. Each channel then updates the list of associated devices if
the device accepts the request and removes the device from the list
of available devices. The request and response phases continue until
all lists of available devices are empty.

\begin{algorithm}[!t]
\begin{algorithmic}[1]
\STATE Creat a list of available devices   for each $\mathrm{CH}_{n},n\in\mathcal{N}$ denoted by $\mathcal{A}_{n}=\mathcal{K},$ and a list of associated devices  for each $\mathrm{CH}_{n},n\in\mathcal{N}$ denoted by $\mathcal{K}_{n}=\emptyset.$

\REPEAT
\STATE Each $\mathrm{DVC}_{k},k\in\mathcal{K}$ sets its perference list $\mathcal{L}_{\mathrm{DVC}_{k}}$ according to \eqref{eq:P-device}, and each $\mathrm{CH}_{n},n\in\mathcal{N}$ sets its perference list $\mathcal{L}_{\mathrm{CH}_{n}}$ according to \eqref{eq:P-MEC}.

\STATE Each $\mathrm{CH}_{n},n\in\mathcal{N}$ proposes itself to its headmost device in $\mathcal{L}_{\mathrm{CH}_{n}}$.

\STATE Each $\mathrm{DVC}_{k},k\in\mathcal{K}$ selects the most preferable channel in $\mathcal{L}_{\mathrm{DVC}_{k}}$ and rejects all the other channels.

\STATE Each $\mathrm{CH}_{n},n\in\mathcal{N}$ updates $\mathcal{K}_{n}$ and $\mathcal{A}_{n}$ as follows:  if the proposed device accepts the request, then add the device into the set $\mathcal{K}_{n}$; if one device rejects the previous accepted request, then delete the device from the set $\mathcal{K}_{n}$; delete the proposed device from the set $\mathcal{A}_{n}$.

\UNTIL{All $\mathcal{A}_{n}$'s are empty.}
\STATE Output: A matching between channels and devices $\xi$.
\end{algorithmic}

\caption{Initial matching between channels and devices based on the Gale-Shapley
algorithm.\label{alg2}}
\end{algorithm}

Then, the swap operation is performed to further increase the MCOR
and achieve the exchange stability, as shown in Algorithm \ref{alg3}.
Specifically, a swap blocking pair is iteratively searched and the
swap matching is updated until the two-sided exchange stability is
achieved.

\begin{algorithm}[!t]
\begin{algorithmic}[1]
\STATE Initialize the matching between channels and devices according to Algorithm \ref{alg2} and denote it as $\xi$.
\REPEAT

\STATE Search a pair of devices $(\mathrm{DVC}_{k},\mathrm{DVC}_{k^{'}})$.
\IF{$(\mathrm{DVC}_{k},\mathrm{DVC}_{k^{'}})$ is a swap blocking pair of the current matching $\xi$}
\STATE Update $\xi$ as $\xi_{k}^{k^{'}}$.
\ENDIF

\UNTIL{No swap blocking pair is found.}
\STATE For all $n\in\mathcal{N}$ and  $k\in\mathcal{K},$ set $\beta_{n,k}=1$ if $\mathrm{CH}_n$ matches with  $\mathrm{DVC}_k$,  and set $\beta_{n,k}=0$ otherwise.
\STATE Output: $\boldsymbol{\beta}.$
\end{algorithmic}

\caption{Swap matching between channels and devices.\label{alg3}}
\end{algorithm}

\subsection{MEC Server Allocation Problem}

After allocating channels to devices, we then assign MEC servers to
(channel, device) units. Note that one MEC server can be allocated
to multiple (channel, device) units, as long as the total computing
frequency at the MEC server is enough to support the total computing
frequency required by these devices. Note also that the MCOR of each
(channel, device) unit at each MEC server does not depend on the other
(channel, device) units sharing the same MEC server, since the computing
frequency required by each device is given. Thus, different from the
channel allocation problem, the MEC server allocation problem is a
one-to-many matching problem without externalities.

Similar to Section \ref{subsec:MEC-Server-Allocation}, the preference
of $(\mathrm{CH}_{n},(\mathrm{DVC}_{k},k\in\mathcal{K}_{n}))$ over
$\mathrm{MEC}_{m}$ is formulated as the MCOR among all associated
devices, which is
\begin{equation}
\mathcal{P}_{m,n,\mathcal{K}_{n}}=\frac{t_{m}^{\mathrm{o}}}{T}\min_{k\in\mathcal{K}_{n}}\sum_{i=1}^{2}r_{m,n,k,i},\label{eq:P-device-1}
\end{equation}
where $\mathcal{S}_{m,n,k,i}$ in $r_{m,n,k,i}$ is given by \eqref{eq:Ski-2}.
Note that the preference of each MEC server $\mathrm{MEC}_{m}$ on
$(\mathrm{CH}_{n},(\mathrm{DVC}_{k},k\in\mathcal{K}_{n}))$ is also
given by $\mathcal{P}_{m,n,\mathcal{K}_{n}}.$ Based on the defined
preference in \eqref{eq:P-device-1}, for a given $(\mathrm{CH}_{n},(\mathrm{DVC}_{k},k\in\mathcal{K}_{n}))$,
any two MEC servers $\mathrm{MEC}_{m}$ and $\mathrm{MEC}_{m^{'}}$,
any two matchings $\xi$ and $\xi^{'}$, we have the following relationships:
\begin{align}
 & (\mathrm{MEC}_{m},\xi)\succ_{(\mathrm{CH}_{n},(\mathrm{DVC}_{k},k\in\mathcal{K}_{n}))}(\mathrm{MEC}_{m^{'}},\xi^{'})\nonumber \\
 & \Leftrightarrow\mathcal{P}_{m,n,\mathcal{K}_{n}}(\xi)>\mathcal{P}_{m^{'},n,\mathcal{K}_{n}}(\xi^{'}),\label{eq:MEC-server-p1}
\end{align}
where $\mathcal{P}_{m,n,\mathcal{K}_{n}}(\xi)$ is the preference
of $(\mathrm{CH}_{n},(\mathrm{DVC}_{k},k\in\mathcal{K}_{n}))$ over
$\mathrm{MEC}_{m}$ under the matching $\xi.$ The expression \eqref{eq:MEC-server-p1}
means that $(\mathrm{CH}_{n},(\mathrm{DVC}_{k},k\in\mathcal{K}_{n}))$
prefers $\mathrm{MEC}_{m}$ in $\xi$ to $\mathrm{MEC}_{m^{'}}$ in
$\xi^{'}$, only when the preference $\mathcal{P}_{m,n,\mathcal{K}_{n}}(\xi)$
is larger than the preference $\mathcal{P}_{m^{'},n,\mathcal{K}_{n}}(\xi^{'})$.
Similarly, for a given MEC server $\mathrm{MEC}_{m}$, any two units
$(\mathrm{CH}_{n},(\mathrm{DVC}_{k},k\in\mathcal{K}_{n}))$ and $(\mathrm{CH}_{n^{'}},(\mathrm{DVC}_{k},k\in\mathcal{K}_{n^{'}}))$,
any two matchings $\xi$ and $\xi^{'}$ with $\mathcal{K}_{n}=\xi_{m}$
and $\mathcal{K}_{n^{'}}=\xi_{m}^{'}$, we have
\begin{align}
 & ((\mathrm{CH}_{n},(\mathrm{DVC}_{k},k\in\mathcal{K}_{n})),\xi)\succ_{\mathrm{MEC}_{m}}((\mathrm{CH}_{n^{'}},(\mathrm{DVC}_{k},\nonumber \\
 & k\in\mathcal{K}_{n^{'}})),\xi^{'})\Leftrightarrow\mathcal{P}_{m,n,\mathcal{K}_{n}}(\xi)>\mathcal{P}_{m,n^{'},\mathcal{K}_{n^{'}}}(\xi^{'}).
\end{align}
The above expression indicates that $\mathrm{MEC}_{m}$ prefers $(\mathrm{CH}_{n},(\mathrm{DVC}_{k},k\in\mathcal{K}_{n}))$
in $\xi$ to $(\mathrm{CH}_{n^{'}},(\mathrm{DVC}_{k},k\in\mathcal{K}_{n^{'}}))$
in $\xi^{'},$ only when $\mathcal{P}_{m,n,\mathcal{K}_{n}}(\xi)$
is larger than $\mathcal{P}_{m,n^{'},\mathcal{K}_{n^{'}}}(\xi^{'}),$
i.e., higher MCOR can be achieved from $(\mathrm{CH}_{n},(\mathrm{DVC}_{k},k\in\mathcal{K}_{n}))$
than that from $(\mathrm{CH}_{n^{'}},(\mathrm{DVC}_{k},k\in\mathcal{K}_{n^{'}}))$.

Based on the constructed preference lists, the Gale-Shapley algorithm
is adopted to achieve a stable matching between (channel, device)
units and MEC servers, as shown in Algorithm \ref{alg5}. Specifically,
all the MEC servers simultaneously propose to the most preferred (channel,
device) unit based on the preference lists. Then, each (channel, device)
unit accepts the most preferred MEC server and rejects all the others.
The algorithm terminates when all (channel, device) units are matched
to the MEC servers.

\begin{algorithm}[!t]
\begin{algorithmic}[1]
\STATE Creat a list of available (channel, device) units   for each $\mathrm{MEC}_{m},m\in\mathcal{M}$ denoted by $\mathcal{A}_{m},$ where each device in $\mathcal{A}_{m}$ satisfies $f_{k}\leq F_{m}$, and a list of associated (channel, device) units  for each $\mathrm{MEC}_{m},m\in\mathcal{M}$ denoted by $\mathcal{B}_{m}=\emptyset.$

\REPEAT
\STATE Each $(\mathrm{CH}_{n},(\mathrm{DVC}_{k},k\in\mathcal{K}_{n})),n\in\mathcal{N}$ sets its perference list $\mathcal{L}_{n}$ according to \eqref{eq:P-device-1}, and each $\mathrm{MEC}_{m},m\in\mathcal{M}$ sets its perference list $\mathcal{L}_{\mathrm{MEC}_{m}}$ according to \eqref{eq:P-device-1}.

\STATE Each $\mathrm{MEC}_{m},m\in\mathcal{M}$ proposes itself to its headmost $(\mathrm{CH}_{n},(\mathrm{DVC}_{k},k\in\mathcal{K}_{n}))$ in $\mathcal{L}_{\mathrm{MEC}_{m}}$.

\STATE Each $(\mathrm{CH}_{n},(\mathrm{DVC}_{k},k\in\mathcal{K}_{n}))$ selects the most preferable MEC server in $\mathcal{L}_{n}$ and rejects all the other MEC servers.

\STATE Each $\mathrm{MEC}_{m},m\in\mathcal{M}$ updates $\mathcal{A}_{m}$ and $\mathcal{B}_{m}$ as follows:  if the headmost (channel, device) unit in $\mathcal{L}_{\mathrm{MEC}_{m}}$ accepts the request, then add the (channel, device) unit into the set $\mathcal{B}_{m}$; if one (channel, device) unit rejects the previous accepted request, then delete the (channel, device) unit from the set $\mathcal{B}_{m}$; delete the proposed (channel, device) unit from the set $\mathcal{A}_{m}$; for each device $k$ in $\mathcal{A}_{m}$, if it satisfies $f_{k}> F_{m}-\sum_{k^{'}\in\mathcal{B}_{m}}f_{k^{'}}$, then delete the  (channel, device) unit from the set $\mathcal{A}_{m}$.

\UNTIL{All $\mathcal{A}_{m}$'s are empty.}
\STATE For all $m\in\mathcal{M}$ and  $k\in\mathcal{K},$ set $\alpha_{m,k}=1$ if $\mathrm{MEC}_m$ matches with  $\mathrm{DVC}_k$,  and set $\alpha_{m,k}=0$ otherwise.
\STATE Output: $\boldsymbol{\alpha}.$
\end{algorithmic}

\caption{Stable matching between (channel, device) units and MEC servers based
on the Gale-Shapley algorithm.\label{alg5}}
\end{algorithm}

\subsection{\textcolor{black}{Complexity Analysis}}

\textbf{\textcolor{black}{}}The complexity of Algorithm \ref{alg2}
executed in step 1 of Algorithm \ref{alg3} mainly depends on the
number of matching procedures. Since the number of devices is $K$
and the number of channels is $N,$ the number of matching procedures
in Algorithm \ref{alg2} is in the order $\mathcal{O}(KN)$. Meanwhile,
the complexity of the remaining steps of Algorithm \ref{alg3} mainly
depends on the number of swapping procedures, which is in the order
$\mathcal{O}(K^{2})$. Thus, the complexity of Algorithm \ref{alg3}
for channel allocation is $\mathcal{O}(KN+K^{2})$. In addition, since
the number of matching procedures in Algorithm \ref{alg5} is in the
order $\mathcal{O}(MN)$, the complexity of Algorithm \ref{alg5}
is $\mathcal{O}(MN)$. Therefore, the total complexity of the proposed
algorithm for MEC server allocation and channel allocation is $\mathcal{O}(KN+K^{2}+MN).$

\section{Simulation Results\label{sec:Simulation-Results}}

In this section, we provide simulation results to confirm the effectiveness
of the proposed algorithm. Unless otherwise noted, the system parameters
during simulation are set as follows: The number of MEC servers is
$M=3$, the number of channels is $N=3$ and the number of devices
is $K=9$; A 2D topology is assumed, where both the devices and MEC
servers are randomly distributed around the origin point within the
distance $0.5$ km; the channels are assumed to follow Rayleigh fading,
i.e., the channel gain $h_{m,n,k}$ is an exponentially distributed
random variable with the mean value corresponding to the path loss
specified by a widely used 3GPP model $128.1+37.6\log_{10}(d)$ in
dB \cite{access2010further,reifert2023rate,diamanti2024delay,yang2022sum,yang2021optimization},
where $d$ is the distance in km; In addition, $T=1$ s, $B=1$ MHz,
$\sigma^{2}=-174$ dBm/Hz, $P_{k}=20$ dBm, and $F_{m}=20$ Mbps.
All results are obtained from $100$ independent simulation runs.

\begin{figure}[!t]
\centering\includegraphics[width=0.8\columnwidth]{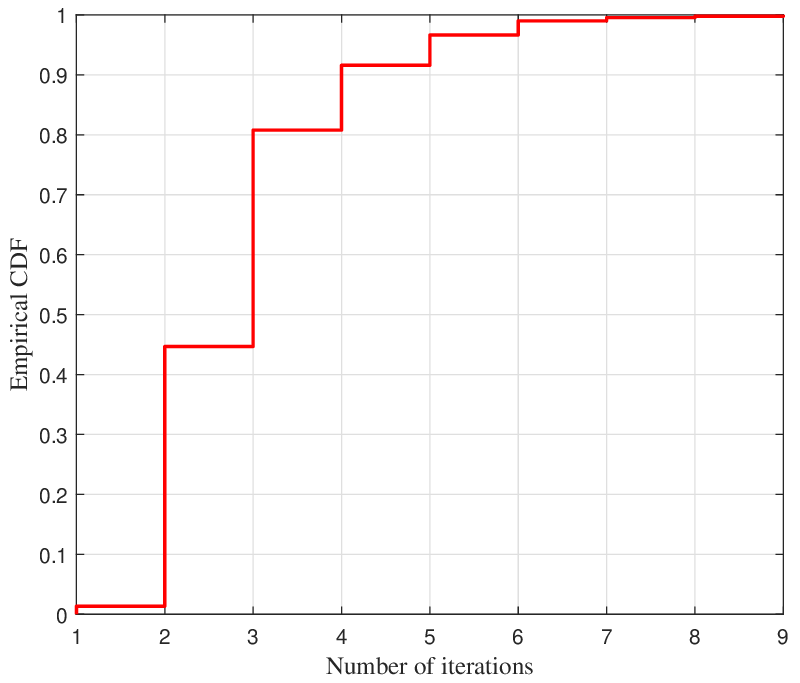}

\caption{CDF of the number of iterations of the SCA method in Algorithm \ref{alg:1}.\label{fig:0-convergence}}
\end{figure}

\begin{figure}[!t]
\centering\includegraphics[width=0.8\columnwidth]{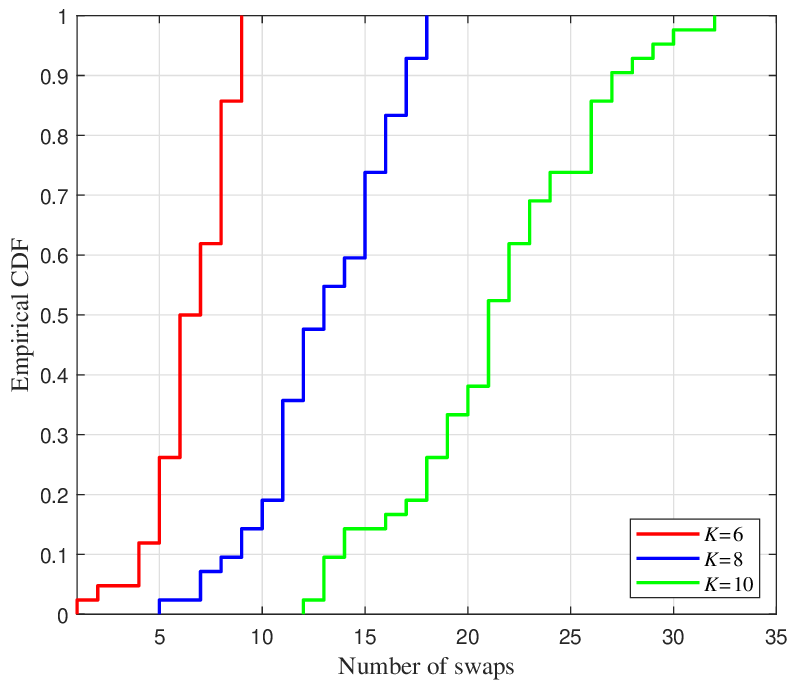}

\caption{CDF of the number of swaps in Algorithm \ref{alg3}.\label{fig:0-convergence3}}
\end{figure}

Fig. \ref{fig:0-convergence} plots the empirical cumulative distribution
function (CDF) of the number of iterations of the SCA method in Algorithm
\ref{alg:1}. It is shown that over $40\%$ and $90\%$ of the iterations
converge within $3$ and $5$ iterations, respectively. These results
indicate that the proposed algorithm is of high efficiency in terms
of convergence speed and can be practically implemented. In Fig. \ref{fig:0-convergence3},
we plot the CDF of the number of swap operations for the swap matching
process in Algorithm \ref{alg3} under different numbers of devices.
It is shown that the proposed swap matching algorithm converges in
only several swap operations. It is also shown that the required number
of swap operations increases as the number of devices increases. This
is due to the fact that a higher number of devices increases the probability
of the existence of swap blocking pairs.
\begin{table}[!t]
\caption{Comparison with the optimal decoding order.\label{tab:Comparison-with-the}}

\centering

\begin{tabular}{|>{\centering}m{1.8cm}|>{\centering}m{0.85cm}|>{\centering}m{0.85cm}|>{\centering}m{0.85cm}|>{\centering}m{0.85cm}|}
\hline 
\multirow{2}{1.8cm}{} & \multicolumn{2}{>{\centering}m{1.7cm}|}{MCOR (Mbps)} & \multicolumn{2}{>{\centering}m{1.7cm}|}{Algorithm execution time (s)}\tabularnewline
\cline{2-5} \cline{3-5} \cline{4-5} \cline{5-5} 
 & $K=2$ & $K=3$ & $K=2$ & $K=3$\tabularnewline
\hline 
\multicolumn{1}{|>{\centering}m{1.8cm}|}{Proposed decoding order} & $2.90$ & $2.28$ & $1.76$ & $5.42$\tabularnewline
\hline 
Optimal decoding order via the exhaustive search & $2.93$ & $2.30$ & $10.04$ & $910.29$\tabularnewline
\hline 
\end{tabular}
\end{table}

\begin{figure}[!t]
\centering\includegraphics[width=0.8\columnwidth]{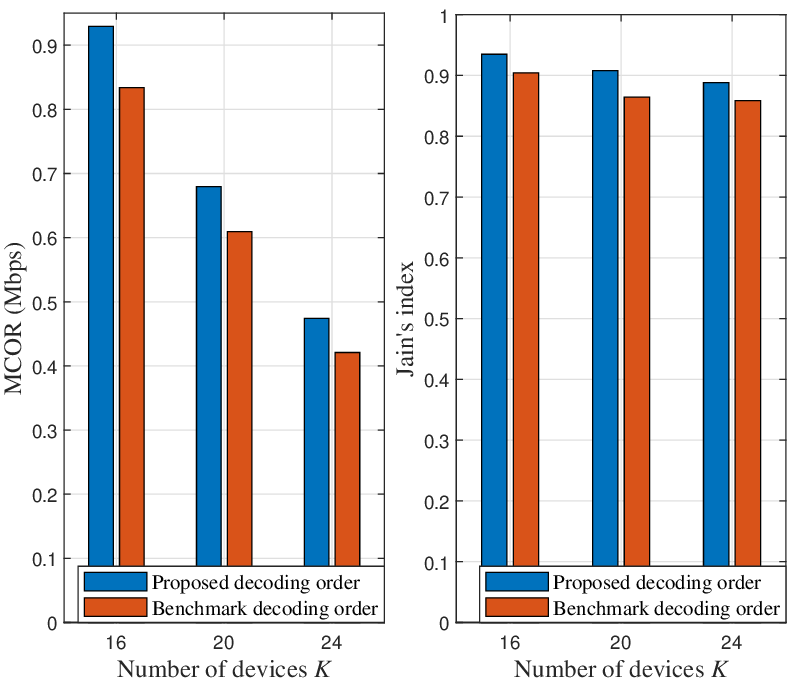}

\caption{Comparison with the benchmark decoding order.\label{fig:0-decoding}}
\end{figure}

In order to show the effectiveness of the proposed decoding order,
Table \ref{tab:Comparison-with-the} compares the proposed decoding
order in Algorithm \ref{alg:1} with the optimal decoding order via
the exhaustive search. Due to the extremely high complexity of the
optimal decoding order, we choose $M=1$, $N=1$, and small numbers
of devices for feasible implementation. It is shown that compared
to the optimal decoding order, the proposed decoding order achieves
slightly lower MCOR with significantly lower algorithm execution time,
especially when the number of devices is relatively large. The results
in Table \ref{tab:Comparison-with-the} show that the proposed decoding
order is effective, i.e., it is of low complexity and close to optimal.

In order to further demonstrate the effectiveness of the proposed
decoding order, in Fig. \ref{fig:0-decoding}, we compare it with
the benchmark decoding order proposed in \cite{yang2022sum} for uplink
RSMA systems under various numbers of devices with $M=2$ and $N=2.$
Specifically, in the benchmark decoding order, the decoding order
of the sub-messages $s_{k,2},\forall k\in\mathcal{K}_{m}$ is the
reverse of the decoding order of the sub-messages $s_{k,1},\forall k\in\mathcal{K}_{m}$.
It is shown that the MCOR achieved by the proposed decoding order
is much higher than the benchmark decoding order. In Fig. \ref{fig:0-decoding},
we also plot the Jain's indexes achieved by different decoding orders.
Note that the Jain's index is a frequently used quantitative fairness
measure, and devices are treated fairer when the Jain's index approaches
to $1$ \cite{huaizhou2014fairness}. It can be seen that the fairness
index decreases slowly as the number of devices increases. This is
because it is generally more harder to make a tradeoff among the performance
of different devices when the number of devices is large. It can be
also seen that the proposed decoding order provides higher fairness
than the benchmark decoding order, and such fairness gain does not
change as the number of devices becomes larger. The results from Fig.
\ref{fig:0-decoding} indicate that the proposed decoding order achieves
not only higher MCOR but also higher fairness than the benchmark decoding
order. 

In what follows, since no current work has investigated the system
model considered in this paper, we design some benchmark algorithms
for comparison with the overall proposed algorithm. In addition, algorithms
designed in existing works are also revised to suit the system model
and are adopted as benchmark algorithms. The benchmark algorithms
are given as follows:
\begin{itemize}
\item RSMA-Random-PropFair: The power allocation and the SIC decoding order
are optimized according to \cite{yang2022sum} with absolute proportional
fairness, the time allocation and the computing frequency allocation
are optimized as given by \eqref{eq:tmo}, \eqref{eq:fmk-2} and \eqref{eq:tmc-2},
and the channels and the MEC servers are randomly allocated. 
\item RSMA-Match-MaxMin: Both the channel allocation and the MEC server
allocation are performed using the matching theory based algorithm
in \cite{zakeri2019joint} to maximize the sum rate, while the remaining
optimization variables are optimized according to the proposed algorithm
to maximize the MCOR. 
\item RSMA-Match-SumRate: In this algorithm, the optimization of the power
allocation and the SIC decoding order is performed to maximize the
sum rate as in \cite{yang2022sum} without considering the proportional
fairness, while the optimization of the remaining variables is the
same as RSMA-Match-MaxMin.
\item RSMA-Random-SumRate: This algorithm is similar to RSMA-Random-PropFair
except that the power allocation and the SIC decoding order are designed
to maximize the sum rate as in \cite{yang2022sum} without considering
the proportional fairness.
\item NOMA-Match: The channel allocation and the MEC server allocation are
performed using the proposed matching theory based algorithm in Section
\ref{sec:MEC-Allocation-and}, the SIC decoding order of all the devices
is in the descending order of channel gain $h_{m,n,k}$ \cite{wang2021sub},
the power allocation is performed according to \cite{kumar2023max},
and the time allocation and the computing frequency allocation are
optimized following the expressions \eqref{eq:tmo}, \eqref{eq:fmk-2}
and \eqref{eq:tmc-2}.
\item NOMA-Random: This algorithm is similar to NOMA-Match except that the
channels and the MEC servers are randomly allocated. 
\item TDMA-Match: The channel allocation and the MEC server allocation are
performed as in Section \ref{sec:MEC-Allocation-and}, while the procedures
for the time allocation and computing frequency allocation are optimized
to maximize the MCOR given in the Appendix. 
\item TDMA-Random: This algorithm is similar to TDMA-Match except that the
channel allocation and the MEC server allocation are randomly performed.
\end{itemize}

\begin{figure}[!t]
\centering\includegraphics[width=0.8\columnwidth]{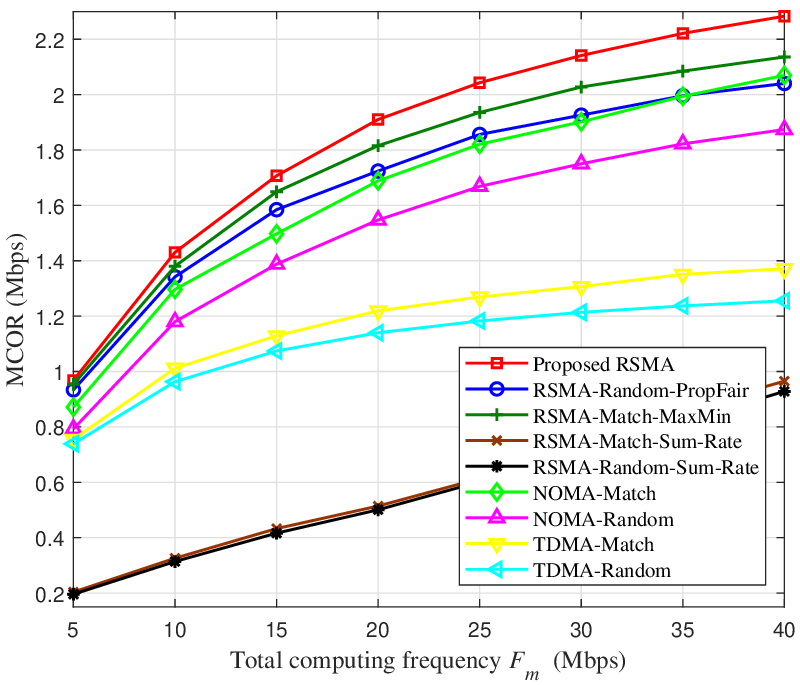}

\caption{MCOR against $\ensuremath{F_{m}}$.\label{fig:1}}
\end{figure}

Fig. \ref{fig:1} shows the MCOR under various values of total computing
frequency $\ensuremath{F_{m}}$ of MEC servers. It is seen that the
MCOR increases as $F_{m}$ increases. This is because a higher computing
frequency can let the devices offload their task data to more proper
MEC servers with a higher computation offloading rate, while a lower
computing frequency leads to offloading data to inferior MEC servers
whose computing frequency may be abundant. Compared to the benchmark
algorithms based on NOMA and TDMA, the proposed algorithm is shown
to achieve a higher MCOR.  It is also shown that the proposed algorithm
outperforms the benchmark algorithms based on RSMA such as RSMA-Random-PropFair
and RSMA-Match-MaxMin, particularly when $\ensuremath{F_{m}}$ is
large. This is because compared to random allocation of MEC servers
in RSMA-Random-PropFair, choosing proper MEC servers by the proposed
algorithm can let more data be computed at the MEC servers, while
compared to random allocation of channels in RSMA-Random-PropFair,
selecting proper channels by the proposed algorithm can let the devices
offload more data to the MEC servers. In addition, the superiority
of the proposed algorithm compared to RSMA-Match-MaxMin has demonstrated
the advantageous of the proposed matching theory based channel and
MEC server allocation algorithm.

From Fig. \ref{fig:1}, RSMA-Match-SumRate and RSMA-Random-SumRate
are shown to significantly underperform other algorithms. This is
because both algorithms aim to maximize the sum rate and do not consider
the device fairness, and this will lead to the devices with inferior
channel conditions being allocated with fewer resources and the devices
with superior channel conditions being allocated with more resources
such that the sum rate can be improved. In addition, it is shown that
the superiority of the proposed algorithm compared to the benchmark
algorithms is more obvious when $\ensuremath{F_{m}}$ is larger. 
This indicates that the proposed algorithm is preferred when the total
computing frequency at the MEC servers is large. 

\begin{figure}[!t]
\centering\includegraphics[width=0.8\columnwidth]{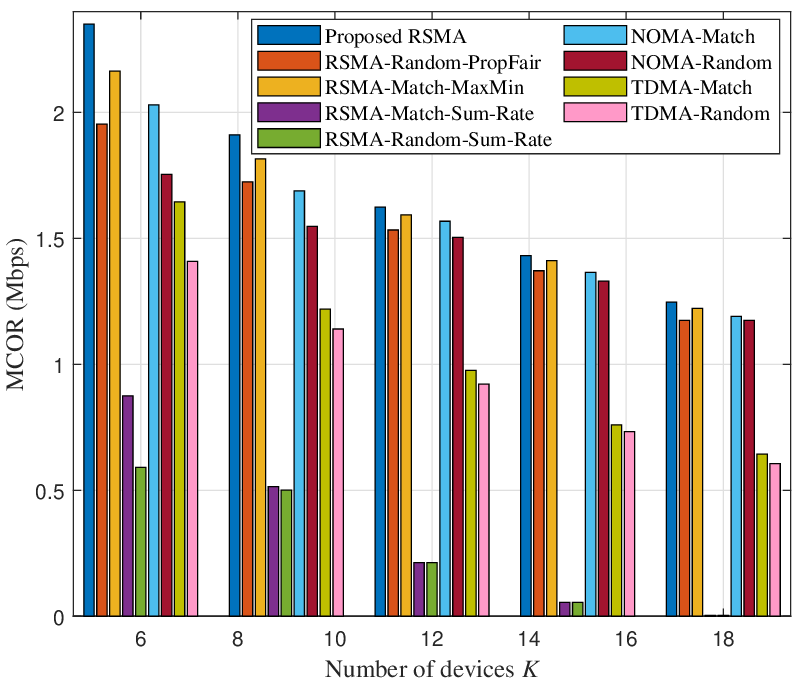}

\caption{MCOR against $K$.\label{fig:2}}
\end{figure}

\begin{figure}[!t]
\centering\includegraphics[width=0.8\columnwidth]{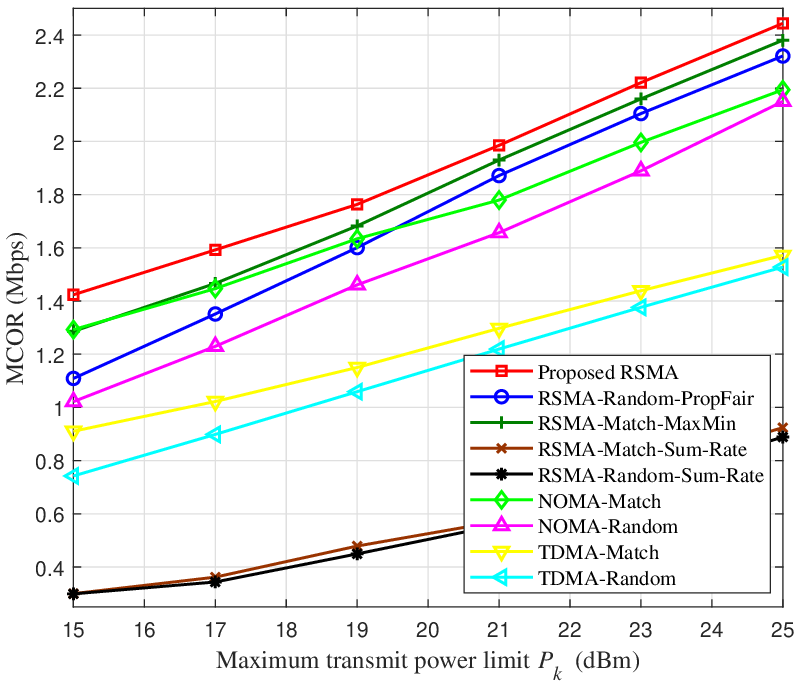}

\caption{MCOR against $P_{k}$.\label{fig:3}}
\end{figure}

Fig. \ref{fig:2} shows the MCOR under various values of the number
of devices $K$. It is seen that as $K$ increases, the MCOR decreases.
The reasons are mainly three-fold. First, more devices will lead to
a higher chance of having a device with inferior channel conditions.
Second, the interference among devices is higher with more devices
sharing the same channel. Third, each device will be allocated with
a lower computing frequency when more devices offload their task data
to the same MEC server.  It is interesting to see that both RSMA-Match-SumRate
and RSMA-Random-SumRate achieve almost zero MCOR when $K$ is very
large. This is due to the fact that a higher $K$ will lead to a higher
probability of having a device with very poor channel conditions such
that almost no resource will be allocated to it if fairness is not
considered. It is also shown that as $K$ increases, the differences
between the proposed algorithm and RSMA-Random-PropFair/RSMA-Match-MaxMin,
between NOMA-Match and NOMA-Random, and between TDMA-Match and TDMA-Random
get smaller. This is because when $K$ increases, it is more likely
to have a device with too poor channel conditions such that its computation
offloading rate cannot be improved no matter which channel or which
MEC server is allocated.

Furthermore, it is shown from Fig. \ref{fig:2} that the proposed
algorithm outperforms NOMA-Match, and the performance gap is obvious
when $K$ is small while it is small when $K$ is large. This phenomenon
is consistent with findings in existing works such as \cite{li2022full}.
The reason for this may be that when the number of devices is large,
it is highly likely to have more devices with poor channel conditions
such that splitting the messages of these devices for controlling
the interference among them will be ineffective in improving the MCOR.
Besides, the algorithms based on RSMA and NOMA outperform the algorithms
based on TDMA and the improvement is more obvious when $K$ is larger.
 The results from Fig. \ref{fig:2} imply that the proposed algorithm
is particularly preferred when the number of devices is small. It
is worth noting that in practical scenarios, due to the complex SIC
operation, the number of devices sharing the same orthogonal channel
in RSMA or NOMA is usually small, which indicates that the proposed
algorithm is useful for practical implementation.

Fig. \ref{fig:3} shows the MCOR under various values of maximum transmit
power of devices $P_{k}$. It is seen that the MCOR increases as $P_{k}$
increases. It is also seen that with the increase of $P_{k}$, the
differences between the proposed algorithm and RSMA-Random-PropFair/RSMA-Match-MaxMin,
between NOMA-Match and NOMA-Random, and between TDMA-Match and TDMA-Random
decrease. This may be because in the higher transmit power region,
the interference among devices gets worse and the system performance
is restricted more by the interference among devices, while the effects
of selecting proper channels and MEC servers get weaker. Furthermore,
it is seen the proposed algorithm outperforms all the benchmark algorithms,
and such performance improvement always holds when $P_{k}$ varies.
Particularly, it is seen that the differences between the proposed
algorithm and NOMA-Match, and between RSMA-Random-PropFair and NOMA-Random
are relatively more obvious when $P_{k}$ is larger.   It is noted
that although RSMA works better than NOMA especially when $P_{k}$
is large, the improvement of RSMA under small $P_{k}$ is still acceptable. 

\begin{figure}[!t]
\centering\includegraphics[width=0.8\columnwidth]{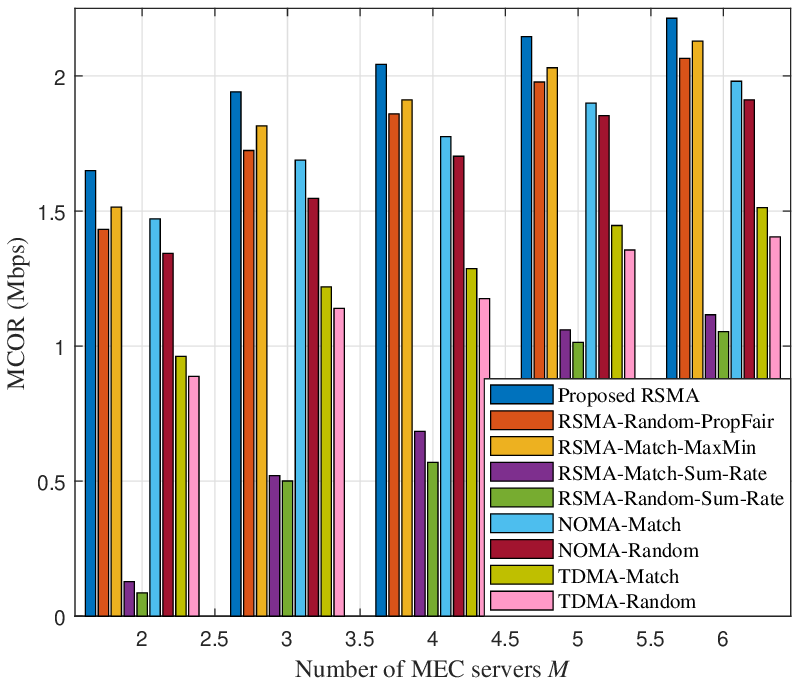}

\caption{MCOR against $M$.\label{fig:4}}
\end{figure}

\begin{figure}[!t]
\centering\includegraphics[width=0.8\columnwidth]{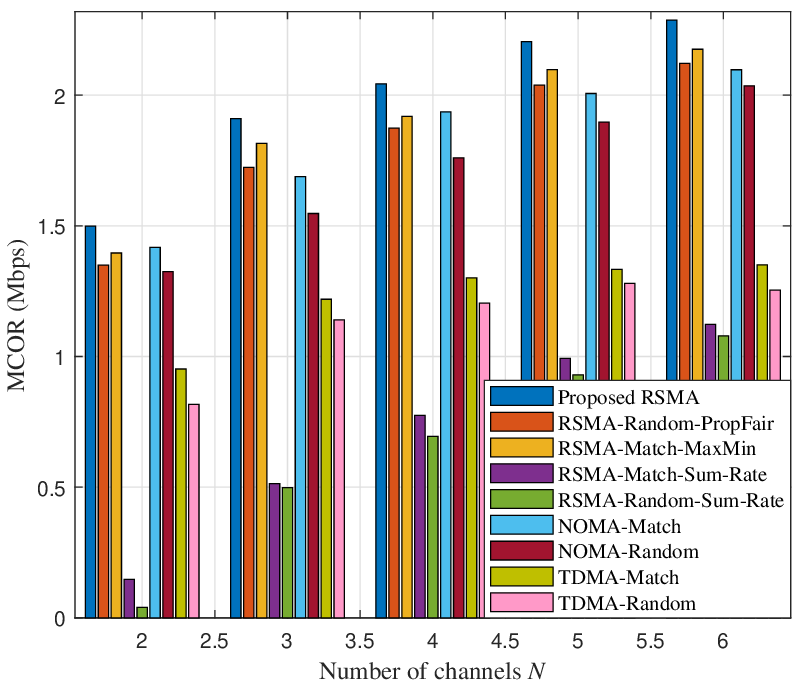}

\caption{MCOR against $N$.\label{fig:5}}
\end{figure}

Fig. \ref{fig:4} shows the MCOR under various values of the number
of MEC servers $M$. It is seen that as $M$ increases, the MCOR increases.
This is due to the fact that more MEC servers can let the devices
choose more proper MEC servers with better channel conditions to
achieve a higher computation offloading rate. In addition, the computing
frequency allocated to each device may also be increased with a larger
$M$, since the number of devices allocated to each MEC server may
be decreased. It is also seen that the performance improvement caused
by increasing $M$ decreases as $M$ increases. This is because when
$M$ is relatively large, as $M$ increases further, the probability
of having an MEC server with better channel conditions is relatively
low. In addition, it is shown that as $M$ increases, the algorithms
based on RSMA always achieve higher MCOR than the algorithms based
on NOMA. This implies that RSMA works well under various values of
the number of MEC servers.

Fig. \ref{fig:5} shows the MCOR under various values of the number
of channels $N$. It is seen that the MCOR increases as $N$ increases.
The reasons for this are mainly three-fold. First, the number of devices
within each RSMA or NOMA group can be decreased with a larger $N$
such that the interference among devices can be reduced. Second, more
channels can let the devices have more chances to experience better
channel conditions. Third, more channels lead to more RSMA or NOMA
groups such that the computing frequency can be more flexibly allocated
to each device. It is also seen that the increase of the MCOR due
to the increase of $N$ slows down as $N$ increases further.  Furthermore,
it is seen that the MCOR achieved by the algorithms based on RSMA
is higher than that achieved by the algorithms based on NOMA and TDMA,
and the rate improvement does not change as $N$ increases. This indicates
that the proposed algorithm based on RSMA is effective under various
values of the number of channels.

\section{Conclusions\label{sec:Conclusions}}

In this paper, we investigate a MCOR maximization problem in an RSMA-assisted
MEC system with multiple MEC servers, multiple channels and multiple
IoT devices, where the channel allocation, the MEC server allocation,
the time allocation, the computing frequency allocation, the transmit
power allocation and the SIC decoding order are jointly optimized.
To solve the problem, an efficient algorithm is designed to achieve
a suboptimal solution. Particularly, given the channel allocation
and the MEC server allocation, the time allocation and the computing
frequency allocation are derived as closed-form functions of the transmit
power allocation and the SIC decoding order, while the transmit power
allocation and the SIC decoding order are optimized based on the alternating
optimization method, the bisection search method and the SCA method.
Then, we construct a hypergraph matching problem to model the channel
and MEC server allocation, and propose a low-complexity matching theory
based algorithm.

The superiority of the proposed algorithm is verified by conducting
simulations. It is demonstrated by simulation results that the proposed
RSMA-assisted MEC system outperforms current MEC systems such as the
NOMA and TDMA assisted MEC systems as well as the benchmark RSMA-assisted
MEC systems, under various system setups. In particular, the impacts
of some important system parameters such as the computing frequency
of MEC servers, the maximum transmit power of devices, the number
of devices, the number of MEC servers and the number of channels on
the system performance are investigated thoroughly. We hope that our
work can serve as a valuable reference and provide meaningful insights
for the further theoretical, algorithmic, and systematic design of
RSMA-assisted MEC systems.

\appendix{}

\setcounter{equation}{0}\renewcommand{\theequation}{A.\arabic{equation}}In
the appendix, the time allocation and computing frequency allocation
in TDMA-Match benchmark algorithm are described. In TDMA-Match, each
device is assigned a fraction of time $t_{m,k}^{\mathrm{o}}$ in $\mathrm{MEC}_{m}$
with transmit power $P_{k}$ for offloading, and the problem of optimizing
the time allocation and computing frequency allocation is formulated
as
\begin{subequations}
\label{eq:tdma}
\begin{alignat}{1}
\max & \:\min_{k\in\mathcal{K}}r_{k}\\
\mathrm{s.t.} & \:\frac{r_{k}T}{f_{m,k}}\leq t_{m,c},\forall m,\forall k\in\mathcal{K}_{m},\label{eq:tdma-c3}\\
 & \:\sum_{k\in\mathcal{K}_{m}}t_{m,k}^{\mathrm{o}}+t_{m}^{\mathrm{c}}\leq T,\forall m,\label{eq:tdma-c4}\\
 & \:t_{m,k}^{\mathrm{o}}\geq0,t_{m}^{\mathrm{c}}\geq0,\forall m,\forall k\in\mathcal{K}_{m},\\
 & \:\eqref{eq:p1-c4},\eqref{eq:p1-c10},\nonumber \\
\mathrm{o.v.} & \:\mathbf{t}^{\mathrm{o}},\mathbf{t}_{c},\mathbf{f}_{m},
\end{alignat}
\end{subequations}
where $\mathbf{t}^{\mathrm{o}}=\{t_{m,k}^{\mathrm{o}},\forall m,\forall k\in\mathcal{K}_{m}\}$,
$r_{k}=\frac{t_{m,k}^{\mathrm{o}}}{T}\sum_{n\in\mathcal{N}}\beta_{n,k}r_{m,n,k}$
and $r_{m,n,k}=B\log_{2}\left(1+\frac{h_{m,n,k}P_{k}}{\sigma^{2}B}\right),$
$\forall m,\forall k\in\mathcal{K}_{m}.$ Similar to Proposition 1
and Proposition 2, the constraints in \eqref{eq:tdma-c3} and \eqref{eq:tdma-c4}
are satisfied with equality by the optimal solution, and thus we can
have
\begin{align}
 & f_{m,k}=\frac{r_{k}T}{T-\sum_{k\in\mathcal{K}_{m}}t_{m,k}^{\mathrm{o}}},\forall m,\forall k\in\mathcal{K}_{m},\label{eq:fmk-1}\\
 & t_{m}^{\mathrm{c}}=T-\sum_{k\in\mathcal{K}_{m}}t_{m,k}^{\mathrm{o}},\forall m.\label{eq:tmc-1}
\end{align}
By inserting the above expressions into the problem \eqref{eq:tdma},
we have
\begin{subequations}
\label{eq:tdma-1}
\begin{alignat}{1}
\max_{\mathbf{t}^{\mathrm{o}}} & \:\min_{k\in\mathcal{K}_{m},m\in\mathcal{M}}\frac{t_{m,k}^{\mathrm{o}}}{T}\sum_{n\in\mathcal{N}}\beta_{n,k}r_{m,n,k}\\
\mathrm{s.t.} & \:\sum_{k\in\mathcal{K}_{m}}t_{m,k}^{\mathrm{o}}\left(\!F_{m}\!+\!\sum_{n\in\mathcal{N}}\beta_{n,k}r_{m,n,k}\!\right)\leq F_{m}T,\forall m,\label{eq:tdma-1-c1}\\
 & \:\sum_{k\in\mathcal{K}_{m}}t_{m,k}^{\mathrm{o}}\leq T,\forall m.\label{eq:tdma-1-c2}
\end{alignat}
\end{subequations}
Further by introducing auxiliary variable $\theta,$ the above problem
can be reformulated as
\begin{subequations}
\label{eq:tdma-2}
\begin{alignat}{1}
\max_{\theta,\mathbf{t}^{\mathrm{o}}} & \:\theta\\
\mathrm{s.t.} & \:\frac{t_{m,k}^{\mathrm{o}}}{T}\sum_{n\in\mathcal{N}}\beta_{n,k}r_{m,n,k}\geq\theta,\forall m,\forall k\in\mathcal{K}_{m},\\
 & \:\eqref{eq:tdma-1-c1},\eqref{eq:tdma-1-c2}.\nonumber 
\end{alignat}
\end{subequations}
 In the above problem, $\theta$ can be optimally obtained by the
bisection search method, and in each search the problem with given
$\theta$ is a linear programming problem and can be optimally solved
via the simplex method \cite{convexop2004}.

\bibliographystyle{IEEEtran}
\bibliography{bibfile}

\begin{thebibliography}{10}
\providecommand{\url}[1]{#1}
\csname url@samestyle\endcsname
\providecommand{\newblock}{\relax}
\providecommand{\bibinfo}[2]{#2}
\providecommand{\BIBentrySTDinterwordspacing}{\spaceskip=0pt\relax}
\providecommand{\BIBentryALTinterwordstretchfactor}{4}
\providecommand{\BIBentryALTinterwordspacing}{\spaceskip=\fontdimen2\font plus
\BIBentryALTinterwordstretchfactor\fontdimen3\font minus
  \fontdimen4\font\relax}
\providecommand{\BIBforeignlanguage}[2]{{%
\expandafter\ifx\csname l@#1\endcsname\relax
\typeout{** WARNING: IEEEtran.bst: No hyphenation pattern has been}%
\typeout{** loaded for the language `#1'. Using the pattern for}%
\typeout{** the default language instead.}%
\else
\language=\csname l@#1\endcsname
\fi
#2}}
\providecommand{\BIBdecl}{\relax}
\BIBdecl

\bibitem{javed2018internet}
F.~Javed, M.~K. Afzal, M.~Sharif, and B.-S. Kim, ``Internet of things ({IoT})
  operating systems support, networking technologies, applications, and
  challenges: A comparative review,'' \emph{IEEE Commun. Surveys Tuts.},
  vol.~20, no.~3, pp. 2062--2100, 2018.

\bibitem{abbas2018mobile}
N.~Abbas, Y.~Zhang, A.~Taherkordi, and T.~Skeie, ``Mobile edge computing: A
  survey,'' \emph{IEEE Internet Things J}, vol.~5, no.~1, pp. 450--465, 2018.

\bibitem{xu2021sum}
D.~Xu and H.~Zhu, ``Sum-rate maximization of wireless powered primary users for
  cooperative {CRNs}: {NOMA} or {TDMA} at cognitive users?'' \emph{IEEE Trans.
  Commun.}, vol.~69, no.~7, pp. 4862--4876, 2021.

\bibitem{zheng2020achievable}
H.~Zheng, K.~Xiong, P.~Fan, Z.~Zhong, Z.~Ding, and K.~B. Letaief, ``Achievable
  computation rate in {NOMA}-based wireless-powered networks assisted by
  multiple fog servers,'' \emph{IEEE Internet Things J.}, vol.~8, no.~6, pp.
  4802--4815, 2021.

\bibitem{mao2022rate}
Y.~Mao, O.~Dizdar, B.~Clerckx, R.~Schober, P.~Popovski, and H.~V. Poor,
  ``Rate-splitting multiple access: Fundamentals, survey, and future research
  trends,'' \emph{IEEE Commun. Surveys Tuts.}, vol.~24, no.~4, pp. 2073--2126,
  2022.

\bibitem{gan2023delay}
Q.~Gan, G.~Li, W.~He, Y.~Zhao, Y.~Song, and C.~Xu, ``Delay-minimization
  offloading scheme in multi-server {MEC} networks,'' \emph{IEEE Wireless
  Commun. Lett.}, vol.~12, no.~6, pp. 1071--1075, 2023.

\bibitem{wang2023joint}
M.~Wang, S.~Shi, D.~Zhang, C.~Wu, and Y.~Wang, ``Joint computation offloading
  and resource allocation for {MIMO-NOMA} assisted multi-user mec systems,''
  \emph{IEEE Trans. Commun.}, vol.~71, no.~7, pp. 4360--4376, 2023.

\bibitem{liu2020resource}
B.~Liu, C.~Liu, M.~Peng, Y.~Liu, and S.~Yan, ``Resource allocation for
  non-orthogonal multiple access-enabled fog radio access networks,''
  \emph{IEEE Trans. Wireless Commun.}, vol.~19, no.~6, pp. 3867--3878, 2020.

\bibitem{xu2024device}
D.~Xu, ``Device scheduling and computation offloading in mobile edge computing
  networks: A novel {NOMA} scheme,'' \emph{IEEE Trans. Veh. Technol.}, vol.~73,
  no.~6, pp. 9071--9076, 2024.

\bibitem{chen2024performance}
B.~Chen and D.~Xu, ``Performance analysis of {NOMA}-based {MEC} systems with
  semi-grant-free transmission,'' \emph{IEEE Wireless Commun. Lett.}, vol.~13,
  no.~4, pp. 1123--1127, 2024.

\bibitem{yu2023computation}
X.~Yu, F.~Xu, J.~Cai, X.-y. Dang, and K.~Wang, ``Computation efficiency
  optimization for millimeter-wave mobile edge computing networks with
  {NOMA},'' \emph{IEEE Trans. Mobile Comput.}, vol.~22, no.~8, pp. 4578--4593,
  2023.

\bibitem{fang2020optimal}
F.~Fang, Y.~Xu, Z.~Ding, C.~Shen, M.~Peng, and G.~K. Karagiannidis, ``Optimal
  resource allocation for delay minimization in {NOMA-MEC} networks,''
  \emph{IEEE Trans. Commun.}, vol.~68, no.~12, pp. 7867--7881, 2020.

\bibitem{kumar2023max}
V.~Kumar, M.~F. Hanif, M.~Juntti, and L.-N. Tran, ``A {Max-Min} task offloading
  algorithm for mobile edge computing using non-orthogonal multiple access,''
  \emph{IEEE Trans. Veh. Technol.}, vol.~72, no.~9, pp. 12\,332--12\,337, 2023.

\bibitem{yang2021optimization}
Z.~Yang, M.~Chen, W.~Saad, and M.~Shikh-Bahaei, ``Optimization of rate
  allocation and power control for rate splitting multiple access ({RSMA}),''
  \emph{IEEE Trans. Commun.}, vol.~69, no.~9, pp. 5988--6002, 2021.

\bibitem{zhou2022rate}
G.~Zhou, Y.~Mao, and B.~Clerckx, ``Rate-splitting multiple access for
  multi-antenna downlink communication systems: Spectral and energy efficiency
  tradeoff,'' \emph{IEEE Trans. Wireless Commun.}, vol.~21, no.~7, pp.
  4816--4828, 2022.

\bibitem{park2023rate}
J.~Park, J.~Choi, N.~Lee, W.~Shin, and H.~V. Poor, ``Rate-splitting multiple
  access for downlink {MIMO}: A generalized power iteration approach,''
  \emph{IEEE Trans. Wireless Commun.}, vol.~22, no.~3, pp. 1588--1603, 2023.

\bibitem{van2023evolutionary}
N.~T.~T. Van, N.~C. Luong, S.~Feng, V.-D. Nguyen, and D.~I. Kim, ``Evolutionary
  games for dynamic network resource selection in {RSMA}-enabled {6G}
  networks,'' \emph{IEEE J. Sel. Areas Commun.}, vol.~41, no.~5, pp.
  1320--1335, 2023.

\bibitem{tegos2022performance}
S.~A. Tegos, P.~D. Diamantoulakis, and G.~K. Karagiannidis, ``On the
  performance of uplink rate-splitting multiple access,'' \emph{IEEE Commun.
  Lett.}, vol.~26, no.~3, pp. 523--527, 2022.

\bibitem{yang2022sum}
Z.~Yang, M.~Chen, W.~Saad, W.~Xu, and M.~Shikh-Bahaei, ``Sum-rate maximization
  of uplink rate splitting multiple access ({RSMA}) communication,'' \emph{IEEE
  Trans. Mobile Comput.}, vol.~21, no.~07, pp. 2596--2609, 2022.

\bibitem{abbasi2023transmission}
O.~Abbasi and H.~Yanikomeroglu, ``Transmission scheme, detection and power
  allocation for uplink user cooperation with {NOMA} and {RSMA},'' \emph{IEEE
  Trans. Wireless Commun.}, vol.~22, no.~1, pp. 471--485, 2023.

\bibitem{han2020rate}
R.~Han, Y.~Wen, L.~Bai, J.~Liu, and J.~Choi, ``Rate splitting on mobile edge
  computing for {UAV}-aided {IoT} systems,'' \emph{IEEE Trans. Cogn. Commun.
  Netw.}, vol.~6, no.~4, pp. 1193--1203, 2020.

\bibitem{reifert2023rate}
R.-J. Reifert \emph{et~al.}, ``Rate-splitting and common message decoding in
  hybrid cloud/mobile edge computing networks,'' \emph{IEEE J. Sel. Areas
  Commun.}, vol.~41, no.~5, pp. 1566--1583, 2023.

\bibitem{diamanti2024delay}
M.~Diamanti, C.~Pelekis, E.~E. Tsiropoulou, and S.~Papavassiliou, ``Delay
  minimization for rate-splitting multiple access-based multi-server {MEC}
  offloading,'' \emph{IEEE/ACM Trans. Netw.}, vol.~32, no.~2, pp. 1035--1047,
  2024.

\bibitem{liu2022rate}
H.~Liu, Y.~Ye, Z.~Bai, K.~J. Kim, and T.~A. Tsiftsis, ``Rate splitting multiple
  access aided mobile edge computing in cognitive radio networks,'' in
  \emph{Proc. IEEE International Conference on Communications Workshops (ICC
  Workshops)}.\hskip 1em plus 0.5em minus 0.4em\relax IEEE, 2022, pp. 598--603.

\bibitem{chen2023rate}
P.~Chen, H.~Liu, Y.~Ye, L.~Yang, K.~J. Kim, and T.~A. Tsiftsis,
  ``Rate-splitting multiple access aided mobile edge computing with randomly
  deployed users,'' \emph{IEEE J. Sel. Areas Commun.}, vol.~41, no.~5, pp.
  1549--1565, 2023.

\bibitem{zhou2024cost}
T.~Zhou, X.~Zeng, D.~Qin, N.~Jiang, X.~Nie, and C.~Li, ``Cost-aware computation
  offloading and resource allocation in ultra-dense multi-cell, multi-user and
  multi-task {MEC} networks,'' \emph{IEEE Trans. Veh. Technol.}, vol.~73,
  no.~5, pp. 6642--6657, 2024.

\bibitem{huaizhou2014fairness}
S.~Huaizhou, R.~V. Prasad, E.~Onur, and I.~Niemegeers, ``Fairness in wireless
  networks: Issues, measures and challenges,'' \emph{IEEE Commun. Surveys
  Tuts.}, vol.~16, no.~1, pp. 5--24, 2014.

\bibitem{du2018computation}
J.~Du, L.~Zhao, J.~Feng, and X.~Chu, ``Computation offloading and resource
  allocation in mixed fog/cloud computing systems with min-max fairness
  guarantee,'' \emph{IEEE Trans. Commun.}, vol.~66, no.~4, pp. 1594--1608,
  2018.

\bibitem{liu2020max}
J.~Liu, K.~Xiong, D.~W.~K. Ng, P.~Fan, Z.~Zhong, and K.~B. Letaief, ``Max-min
  energy balance in wireless-powered hierarchical fog-cloud computing
  networks,'' \emph{IEEE Trans. Wireless Commun.}, vol.~19, no.~11, pp.
  7064--7080, 2020.

\bibitem{ju2022energy}
H.~Ju, S.~Kim, Y.~Kim, and B.~Shim, ``Energy-efficient ultra-dense network with
  deep reinforcement learning,'' \emph{IEEE Trans. Wireless Commun.}, vol.~21,
  no.~8, pp. 6539--6552, 2022.

\bibitem{pan2024energy}
T.~Pan, X.~Wu, T.~Zhang, and X.~Li, ``Energy-efficient resource allocation in
  ultra-dense networks with {EMBB} and {URLLC} users coexistence,'' \emph{IEEE
  Trans. Veh. Technol.}, vol.~73, no.~2, pp. 2549--2563, 2024.

\bibitem{convexop2004}
S.~Boyd and L.~Vandenberghe, \emph{Convex Optimization}.\hskip 1em plus 0.5em
  minus 0.4em\relax Cambridge, U.K.: Cambridge Univ. Press, 2004.

\bibitem{cvx}
M.~Grant and S.~Boyd, ``{CVX}: Matlab software for disciplined convex
  programming, version 2.1,'' \url{http://cvxr.com/cvx}, Mar. 2014.

\bibitem{zhang2017hypergraph}
H.~Zhang, L.~Song, Y.~Li, and G.~Y. Li, ``Hypergraph theory: Applications in
  {5G} heterogeneous ultra-dense networks,'' \emph{IEEE Commun. Mag.}, vol.~55,
  no.~12, pp. 70--76, 2017.

\bibitem{gale1962college}
D.~Gale and L.~S. Shapley, ``College admissions and the stability of
  marriage,'' \emph{The American Mathematical Monthly}, vol.~69, no.~1, pp.
  9--15, 1962.

\bibitem{access2010further}
``Further advancements for {E-UTRA} physical layer aspects (release 9),'' 3GPP,
  TS 36.814 (V9.0.0), Mar. 2010.

\bibitem{zakeri2019joint}
A.~Zakeri, M.~Moltafet, and N.~Mokari, ``Joint radio resource allocation and
  {SIC} ordering in {NOMA}-based networks using submodularity and matching
  theory,'' \emph{IEEE Trans. Veh. Technol.}, vol.~68, no.~10, pp. 9761--9773,
  2019.

\bibitem{wang2021sub}
K.~Wang, F.~Fang, D.~B. Da~Costa, and Z.~Ding, ``Sub-channel scheduling, task
  assignment, and power allocation for {OMA}-based and {NOMA}-based {MEC}
  systems,'' \emph{IEEE Trans. Commun}, vol.~69, no.~4, pp. 2692--2708, 2021.

\bibitem{li2022full}
T.~Li, H.~Zhang, X.~Zhou, and D.~Yuan, ``Full-duplex cooperative rate-splitting
  for multigroup multicast with {SWIPT},'' \emph{IEEE Trans. Wireless Commun.},
  vol.~21, no.~6, pp. 4379--4393, 2022.

\end{thebibliography}

\end{document}